\documentclass[manuscript]{aastex63}

\submitjournal{The Astrophysical Journal}

\shorttitle{Thermal History Uncertainty Quantification}
\shortauthors{Seales and Lenardic}
\graphicspath{{./}{ApJ_manuscript/Figures/}}

\begin{document}

\title{Uncertainty Quantification in Planetary Thermal History Models: Implications for Hypotheses Discrimination and Habitability Modeling}

\correspondingauthor{Johnny Seales}
\email{jds16@rice.edu}

\author[0000-0002-0786-7307]{Johnny Seales}
\affiliation{Rice University \\
Department of Earth, Environmental and Planetary Science 
6100 Main St.  \\
Houston, TX 77005, USA}

\author{Adrian Lenardic}
\affiliation{Rice University \\
Department of Earth, Environmental and Planetary Science 
6100 Main St.  \\
Houston, TX 77005, USA}

\begin{abstract}

Multiple hypotheses/models have been put forward regarding Earth's cooling history. Searching for life beyond Earth has brought these models into a new light as they connect to an energy source life can tap. Discriminating between different cooling models and adopting them to aid in the assessment of planetary habitability has been hampered by a lack of uncertainty quantification. Here we provide an uncertainty quantification that accounts for a range of interconnected model uncertainties. This involved calculating over a million individual model evolutions to determine uncertainty metrics. Accounting for uncertainties means that model results must be evaluated in a probabilistic sense, even though the underlying models are deterministic. The uncertainty analysis was used to quantify the degree to which different models can satisfy observational constraints on the Earth's cooling. For the Earth's cooling history, uncertainty leads to ambiguity - multiple models, based on different hypotheses, can match observations. This has implications for using such models to forecast conditions for exoplanets that share Earth characteristics but are older than the Earth, i.e., ambiguity has implications for modeling the long-term life potential of terrestrial planets. Even for the most Earth-like planet we know of, the Earth itself, model uncertainty and ambiguity leads to large forecast spreads. Given Earth has the best data constraints, we should expect larger spreads for models of terrestrial planets in general. The uncertainty analysis provided here can be expanded by coupling planetary cooling models to climate models and propagating uncertainty between them to assess habitability from a probabilistic view.

\end{abstract}



\section{Introduction}
The surface conditions of the Earth have evolved over our planet's history in response to two energy sources: solar energy and internal energy. Both energy sources have, themselves, evolved and continue to do so. Stellar models provide insights into the Sun's energetic evolution \citep{Feulner2012}. Thermal history models provide insights into the cooling of the Earth's interior \citep{Davies1980, Schubert1980}. Earth's internal energy comes from the decay of radioactive isotopes within its rocky interior and from heat retained from planetary formation and early differentiation. This internal energy drives volcanic and tectonic activity, both of which influence the cycling of life-essential elements and volatile elements, such as greenhouse gasses, between the Earth's interior and surface reservoirs (atmosphere, hydrosphere, biosphere). That connection to elemental cycling, along with the discovery of life that can tap into the Earth's internal energy \citep{Baross1985,Jannasch1985} and an expanding search for life beyond Earth, has rejuvenated interest in the cooling history of the Earth and, by association, thermal history models. This renaissance has moved thermal history modeling from the realm of geosciences into the realm of astronomy and astrophysics \citep{Kite2009, Schaefer2015, Komacek2016, Foley2015, Foley2016, Tosi2017, Foley2017, Rushby2018, Barnes2020}.

When a modeling methodology moves from one discipline to another there is the potential for synergies and for misconceptions. The Earth has the largest observational data set that can constrain planetary models. However, this does not mean that significant uncertainties do not remain. This has not been communicated as well as it could be across communities. Even within the geosciences's community itself the role of uncertainty and ambiguity for thermal history models has not received the level of attention given to it in other modeling endeavors (e.g., water resources, climate \citep{Loucks, Curry2011}). This provides the two-pronged motivation for this paper: 1) Given data and model uncertainties, what is the confidence level we can give to different Earth cooling models and, by association, are multiple models viable?; 2) What implications does uncertainty regarding the Earth's thermal history carry for modeling the habitability of terrestrial planets?

\begin{figure}[h!]
  \centering
  \includegraphics[width=0.5\linewidth]{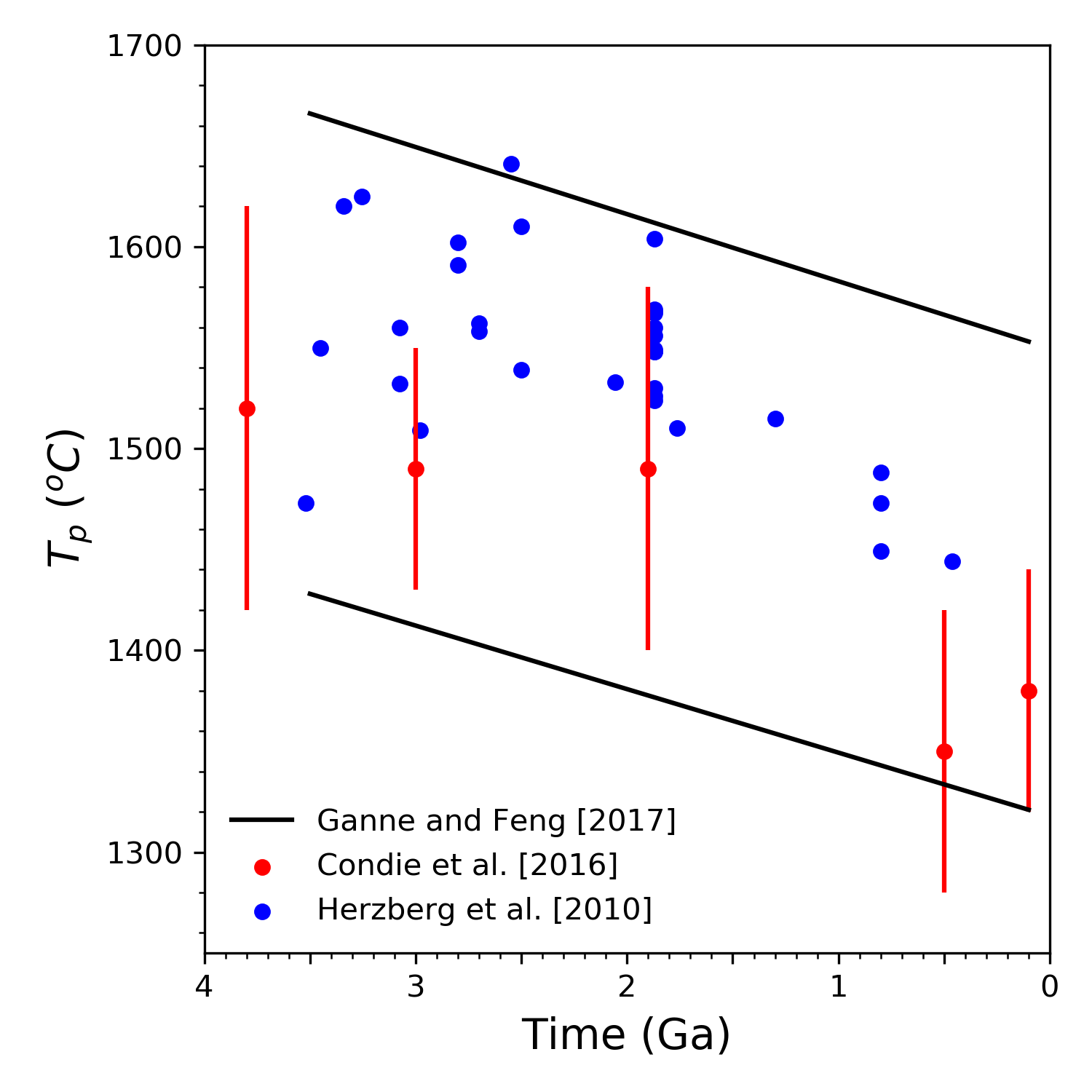}
  \caption{Geologic proxy data for mantle potential temperature ($T_p$) throughout Earth's history. We use \citep{Ganne2017} as a constraint for our model.}
  \label{fig:data}
\end{figure}

The cooling history of a planet depends on its tectonic mode \citep{Lenardic}. The Earth's present mode is plate tectonics \citep{McKenzie1967,Morgan1968}. The simplest starting assumption is that plate tectonics has operated over the Earth's geologic history (i.e., since the transition from a magma ocean phase to a phase of planetary evolution that preserves a rock record \citep{Sleep2000}). This assumption has been made by the majority of Earth thermal history models to date, and we will follow suit herein. With knowledge of our conclusions, we can say that geologic proxy data (\ref{fig:data}) used to constrain the Earth's cooling cannot rule out this possibility. Models that allow for tectonic transitions may also be able to match data constraints, but that will only increase the effects of model uncertainty. By assuming a single tectonic mode we will not only follow an Occam's razor approach, but we will also be conservative in assessing model uncertainty. 

The theory of plate tectonics is a kinematic one that defines the Earth's surface as being divided into internally rigid, rocky plates that move relative to each other with deformation and volcanic activity concentrated along plate boundaries. For the Earth's cooling, a key factor is that cold tectonic plates can sink back (subduct) into the Earth's warmer, rocky interior (i.e., plates are a component of the upper thermal boundary layer of the solid Earth's thermal convection system). Extending the kinematic theory of plate tectonics to a dynamic one involves quantifying the forces that drive and resist plate motions. Which are the primary, or dominant, forces and their magnitude within this force balance remains a debated issue. This debate is central to this paper, as it means that different models, based on different assumptions regarding the forces that resist plate motion, have been proposed.

The range of proposed plate tectonic cooling models for the Earth differ significantly in terms of physical assumptions, and each, therefore represents a different hypotheses regarding the dynamics of plate tectonics. The cooling rate associated with the convective overturn of tectonic plates depends on resisting forces to plate motion. The earliest plate tectonic cooling models assumed that the dominant resistance to plate motions comes from the viscosity of the Earth's mantle - the rocky interior that plates move over and subduct into \citep{Tozer1972,Schubert1979,Schubert1980}. Later models argued that the strength of plates needed to be considered as plate deformation and deformation at plate boundaries provided significant energy dissipation \citep{Conrad1999,Conrad1999b}. Those models assumed that plate strength would decrease under hotter conditions, i.e., in the Earth's past, or remain constant. That assumption was challenged by another plate tectonic cooling model that assumed plate strength increased in the Earth's past \citep{Korenaga2003,Korenaga2006}. All of these models remain argued for to this day (discussed in more detail in the next section) with different authors arguing with variable degrees of 'argumentative force.' The fact that debate remains signals that there is no singular, agreed upon, plate tectonic cooling model, which has implications for modeling planetary habitability beyond Earth. Models that couple interior planet cooling to climate evolution, seeking to address long term habitability of terrestrial planets in general, consider the potential of different tectonic modes, with one example being a plate-tectonic cooling model \citep{Driscoll2013, Foley2016}. A misconception that can follow is that there is a singular, agreed upon, plate tectonic cooling model. As noted above, and detailed in what follows, this is not correct. 

How different are proposed plate tectonic cooling models in effect? That is, are the differences in terms of model outputs small relative to data uncertainty? Over the full range of the models that have been proposed to date, they are not. This is clearly demonstrated in the fact that the sign of the dominant feedback for planetary cooling varies from negative to positive over the full range of proposed models \citep{Moore2015,Seales2019}; the dominant feedback determines whether plate tectonics is less (positive feedback) or more efficient (negative feedback) at cooling the mantle at hotter temperatures. The implications for extrapolating Earth cooling models to "Earth-like" terrestrial planets is significant \citep{Tozer1972,Korenaga2016}. 

To date, no study has systematically compared model outputs for the range of proposed plate tectonic thermal history models to observational data in light of model uncertainties, though some have considered uncertainty in specific contexts \citep{McNamara2000, Korenaga2011}. The bulk of this paper sets out to provide such a comparison. First, the comparison is carried out for model evolutions over the Earth's geologic age. That exercise will isolate models that are consistent with Earth data constraints. From there, we will project this range of ``successful" models forward in time to model Earth-like planets older than the Earth. This will provide insights into the level of certainty that exists for making statements regarding the thermal state of terrestrial planets assumed to operate in a plate tectonic cooling mode, an issue of interest to the planetary habitability community. 

\section{Methods}
In this section we will define the thermal history models we used in this analysis, define the model uncertainties we evaluated, outline the geologic proxy data we used as model constraints, and define how these were combined to assign probabilities of model success. 

In principle, thermal history models can be formulated to solve for the full three dimensional evolution of a planetary interior over time \citep[e.g.,][]{Zhong2000}. In practice, such formulations (run over the Earth's full geological history) remain computationally expensive, which limits the degree to which model output space can be explored. For this reason, thermal history models of the Earth have been formulated to track the average internal temperature of the Earth, and the majority of thermal history models presented for the Earth are of this variety. 

Thermal history models that track averaged internal temperatures are also referred to as parameterized thermal history models. Different parameterizations reflect different assumptions regarding the operation of plate tectonics (discussed more fully below). That difference being noted, thermal history models share a common underpinning: The Earth's average mantle temperature evolves over time based on the balance between heat produced within ($H$) and lost from ($Q$) the mantle according to

\begin{equation}\label{basic}
    C\dot{T_p}=H-Q.
\end{equation}

where the temperature here is the mantle potential temperature ($T_p$). That is the temperature a parcel of mantle with temperature $T_m$ would be if it were brought to the surface adiabatically. Heat is produced within the mantle by the radiogenic decay of $^{238}U$, $^{235}U$, $^{232}Th$ and $^{40}K$, and heat production over time is given by 

\begin{equation}\label{H}
    H(t)=H_0\sum_{n=1}^4h_nexp(\lambda_nt),h_n=\frac{c_np_n}{\sum_nc_np_n}
\end{equation}
where $H_0$ is a reference heat production, $h_n$ is the amount of heat produced by a given isotope, and \emph{t} is time. We calculate relative isotopic concentrations by assuming present day proportions of $U:Th:K=1:4:(1.27x10^4$ and normalizing by total U \citep{Turcotte2002}. The values used in equation \ref{H} are listed in Table \ref{table:radiogenics}.

\begin{table}[h!]
\caption{Radiogenic Heat Production}
\centering
\begin{tabular}{ c c c c c }
\hline\hline
Isotope & $p_n$ $(W/kg)$ & $c_n$ & $h_n$ & $\lambda_n$ $(1/Ga)$ \\
\hline
$^{238}U$  & $9.37\times 10^{-5}$ & 0.9927 & 0.372  & 0.155 \\
$^{235}U$  & $5.69\times 10^{-4}$ & 0.0072 & 0.0164 & 0.985 \\
$^{232}Th$ & $2.69\times 10^{-5}$ & 4.0    & 0.430  & 0.0495 \\
$^{40}K$   & $2.79\times 10^{-5}$ & 1.6256 & 0.181  & 0.555 \\
\hline
\end{tabular}
\label{table:radiogenics}
\end{table}

Heat from the Earth's metallic core could be more directly included in equation \ref{basic} by building in a core evolution model. For simplicity, and to be consistent with the bulk of previous Earth cooling models, we will not do so herein. Adding a core evolution model would only increase model uncertainties, and in not doing so we will follow the approach of being conservative in our uncertainty assessment. We also leave out tidal heating as it is not a major effect in the Earth context. Including it as a heat source term could be an interesting extension of this work for planetary bodies such as those of the TRAPPIST-1 system, but this is outside the scope of our analysis. 

Heat is lost from the planetary interior by convective cooling. This cooling is parameterized according to the Nusselt-Rayleigh scaling law given by

\begin{equation}\label{Nu_Ra}
    Nu\sim Ra^\beta
\end{equation}
\citep{Turcotte2002}. The Nusselt number \emph{Nu} is a nondimensional heat flux calculated as the ratio of convective (\emph{Q}) to conductive (\emph{q}) heat flux across the convecting layer. Conductive heat flux is given by: $q=\frac{k\Delta T}{D}$, where \emph{k}, \emph{D} and $\Delta T$ are the thermal conductivity, convecting layer thickness and temperature drop across the convecting layer, respectively. The Rayleigh number \emph{Ra} is a nondimensional number that describes the vigor of convection and is defined as

\begin{equation}\label{Ra}
    Ra=\frac{\rho g\alpha\Delta TD^3}{\kappa\eta}
\end{equation}

where $\rho$ is mantle density, \emph{g} is the acceleration due to gravity, $\alpha$ is the thermal expansivity, $\Delta T$ is the temperature between the surface and interiork, $\kappa$ is mantle diffusivity and $\eta$ is mantle viscosity. Here we assume surface temperature is zero and thus $\Delta T$ reduces to $T_p$. To change Nusselt-Rayleigh scaling to an equivalency requires a constant \emph{a} be added to the right hand side of equation \ref{Nu_Ra}. The value of \emph{a} is dependent on the geometry of the convecting system and the average aspect ratio of convection cells. One can use laboratory experiments, boundary layer theory, and/or numerical simulations to constrain the constant \citep{Davies1980, Schubert1980}. One may also scale the heat flow to present day heat flow $Q_0$ and employ a scaling temperature $T_0$ as was down by \citet{Christensen1985} and \citet{Korenaga2003} to fix the scaling constant \emph{a} and arrive at a heat flow scaling given by

\begin{equation}\label{HeatFlow}
    Q = Q_0\left(\frac{T_p}{T_0}\right)^{1+\beta}\left(\frac{\eta (T_0)}{\eta (T_p)}\right)^{\beta}.
\end{equation}

Equation \ref{HeatFlow} is dependent on viscosity ($\eta(T_p)$) which is defined as

\begin{equation}\label{eta}
    \eta (T_p)=\eta_0\exp\left(\frac{A}{RT_p}\right)
\end{equation}

where  \emph{A}, \emph{R} and $\eta_0$ are the activation energy, universal gas constant and scaling constant \citep{Karato771}, respectively. For comparison to previous work, we set $\eta_0$ so that the upper mantle has a viscosity of $10^{19}$ Pa$\cdot$ s at 1350 $^oC$. Combining equations \ref{Nu_Ra}-\ref{eta} and using the definition of \emph{Nu} leads to the governing equation 

\begin{equation} \label{Ebal}
C\dot{T_p}=H_0\sum_{n=0}^{4} h_nexp\left(-\lambda_nt\right)-Q_0\left(\frac{T_p}{T_0}\right)^{1+\beta}\left(\frac{\eta (T_0)}{\eta (T_p)}\right)^{\beta}.
\end{equation}

\begin{figure}[h!]
  \includegraphics[width=\linewidth]{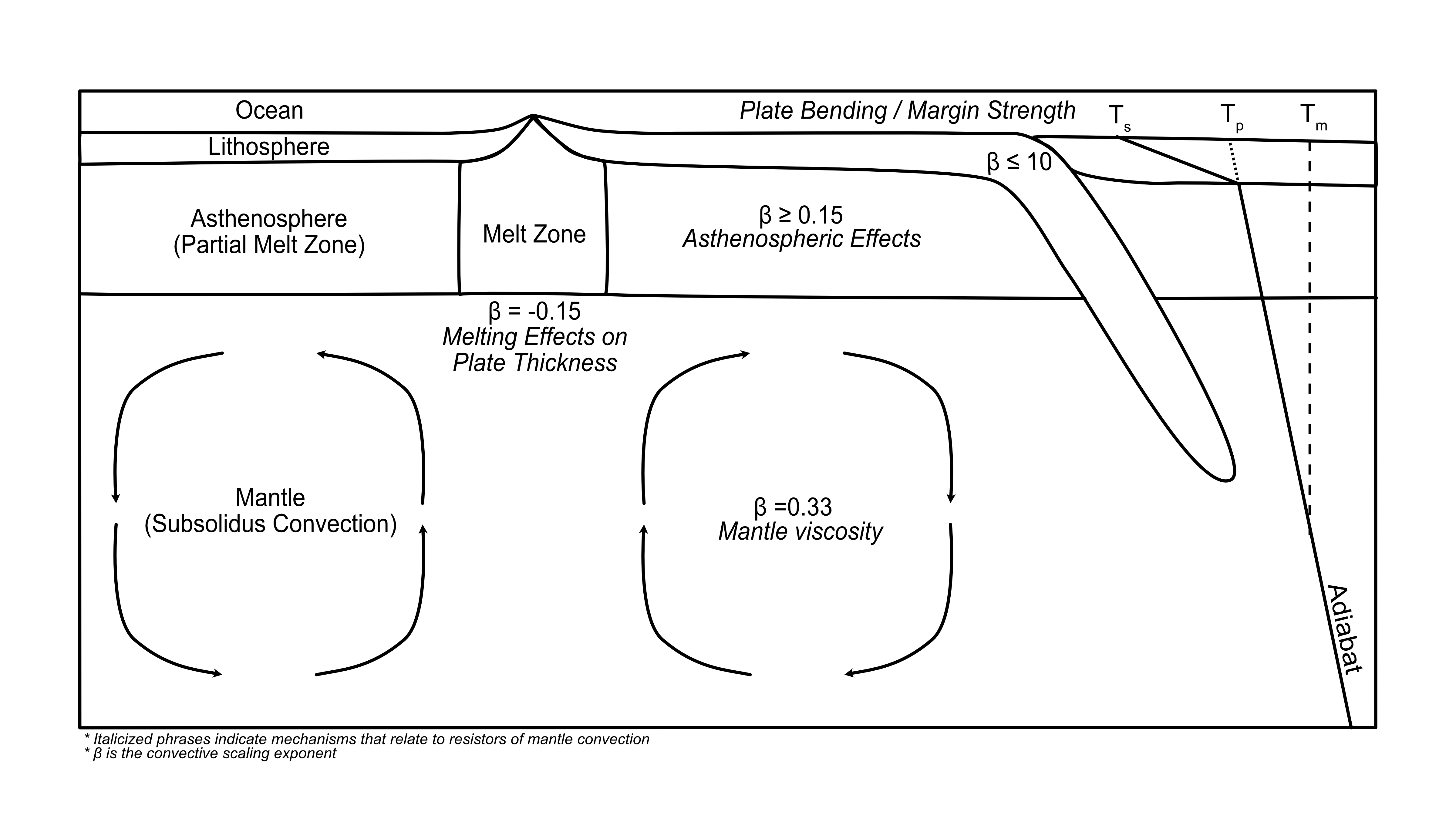}
  \caption{The sources of plate resisting forces, effective $\beta$ value associated with different hypotheses regarding the dynamics of plate tectonics, and a thermal depth profile relating mid mantle temperatures ($T_m$) to mantle potential temperatures ($T_p$).}
  \label{fig:model_cartoon}
\end{figure}

Choosing the value for $\beta$ in equation \ref{Ebal} involves making assumptions/hypotheses regarding the dynamics of plate tectonics (\ref{fig:model_cartoon}). The earliest thermal history models used a value of 0.33 \citep{Schubert1980,Spohn1982,Jackson1984}. This assumes that the dominant resistance to convective motion comes from mantle viscosity \citep{Tozer1972}. It also assumes very vigorous convection. For levels of convection pertinent to the Earth the scaling exponent is slightly lower, $0.30<=\beta<=0.32$ \citep{Schubert1985,Lenardic2003}, due to the upper boundary of mantle convection (i.e., plates in a plate tectonic mode) not being fully self-determined \citep{Moore2015}. Later models, that more directly incorporated model analogues to tectonic plates, showed that values nearly matching this scaling would be recovered provided that very weak plate boundaries were also incorporated \citep{Gurnis1989}. Later models that allowed weak plate boundaries to develop dynamically lead to a scaling exponent of 0.29 \citep{Moresi1998}. If plate boundaries are not assumed to be so weak that energy dissipation along them can be neglected and/or if plate strength offers significant resistance, then the scaling exponent will be significantly lower with a range between $0<=\beta<=0.15$ having been proposed \citep{Christensen1985,Giannandrea1993,Conrad1999b,Conrad1999}. A low viscosity channel below plates - the Earth's asthenosphere \citep{Richards2018} - allows different size plates to have different balances between plate driving and resisting forces \citep{Hoink2011}. This leads to a mixed mode scaling in which plate strength is the dominant resistance for small plates while mantle viscosity is dominant for larger plates. Considering the distribution of current tectonic plate sizes as a guide, this leads to a global heat flow scaling exponent of $0.15<=\beta<=0.20$ \citep{Hoink2013}. An argument for $\beta<0$ has also been made \citep{Korenaga2003}. The physical basis for this last class of models is that at hotter mantle temperatures enhanced melting would generate a thicker dehydrated layer below oceanic crust. This layer would be responsible for the bulk of plate strength. By this reasoning, hotter mantle temperatures in Earth's past would allow for a thicker, stronger plates, which would slow plate velocities and decrease the rate at which the mantle cooled. 

Given that different $\beta$ values represent different physical assumptions regarding the dynamics of plate tectonics, and by association Earth cooling, it follows that different values of $\beta$ represent different hypotheses. This means that we can think of the choice of $\beta$ as a model selection problem, which introduces model selection uncertainty into our analysis. To account for this, we will assume the different models historically put forth are unique; however, we will allow for $\beta$ values between them to represent gradational changes between the different hypotheses. Specifically, we will test a range of models with $\beta$ values between -0.15 and 0.3 at intervals of 0.025. In doing so, our analysis will generate relatively smooth model probability distributions in $\beta$ space, allowing us to map peaks in $\beta$ space to determine models with the highest probability of matching data constraints subject to a variety of uncertainties.

For each $\beta$ model, we will also evaluate combined initial condition and parametric uncertainty. The values for each are listed in Table \ref{table:convectiion_parameters}. Initial condition uncertainty for thermal history models comes from uncertainties about post-magma-ocean planetary temperatures. Parametric uncertainty for thermal history models is connected to the values used for radiogenic heating, the heat flow scaling constant, and the scaling temperature. The strength of temperature dependent viscosity is also a model parameter that can be subjected to a range of values. For simplicity, we will not consider that explicitly herein, as it is connected to variations in the mantle Rayleigh number, \emph{Ra}, which will already be subjected to a range of variations due to the variations in the other parameters noted. 

\begin{figure}[h!]
  \includegraphics[width=\linewidth]{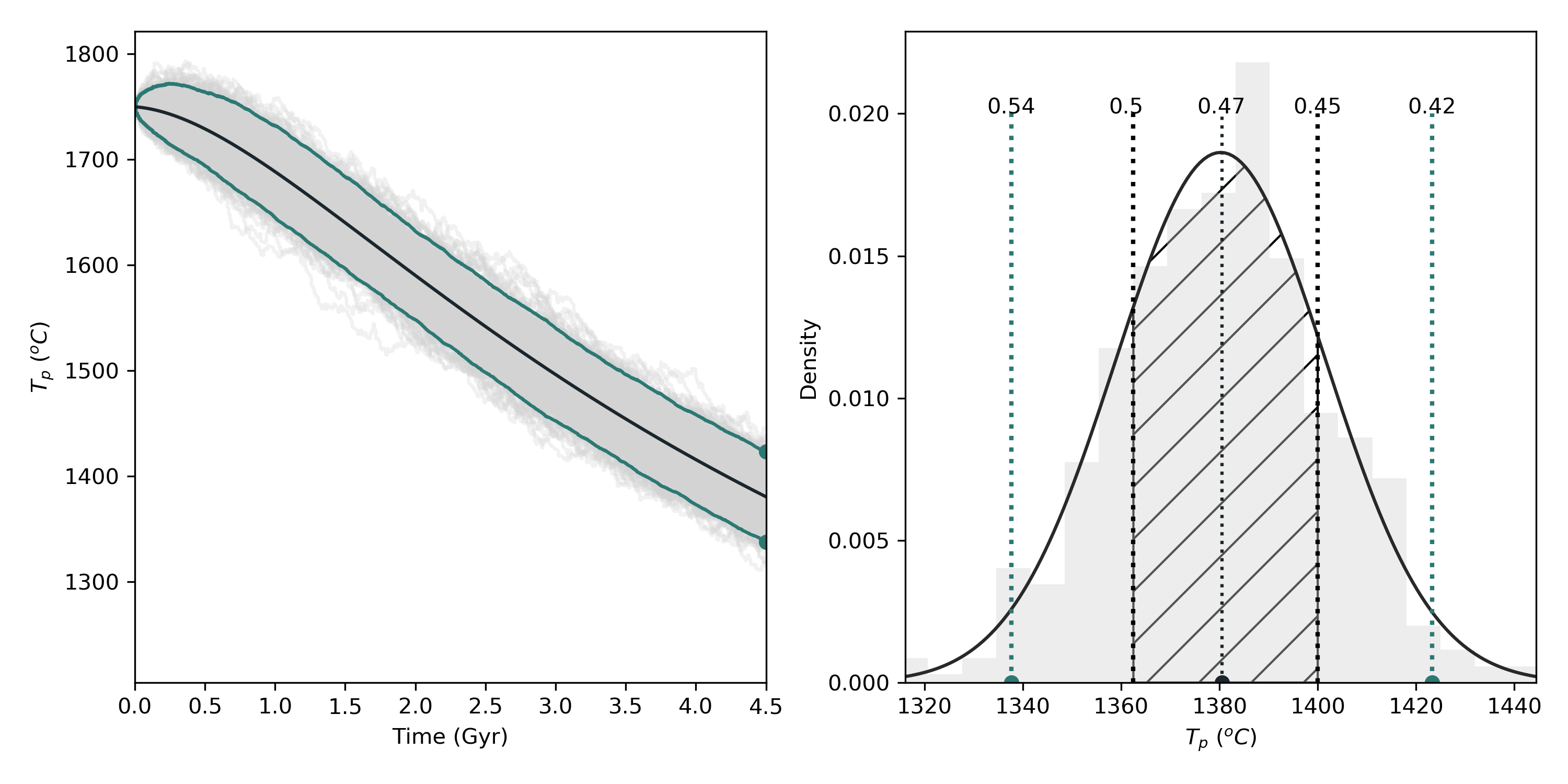}
  \caption{Ensemble model runs for a single thermal history model (a) and an uncertainty measure for a single ensemble (b). An ensemble of 100 perturbed paths (gray lines) is plotted in (a), along with the ensemble mean (blue line) and ensemble two standard deviation limits (red lines). In (b) a present day time slice is taken through the ensemble evolution. This ensemble has an approximately 50\% probability of satisfying present day constraints, which are temperature (1300$^oC$-1400$^oC$) and a Urey ratio between 0.2 and 0.5. Urey ratios are listed are labeled for the the mean, uncertainty limit and acceptable bound solutions. }
  \label{fig:sample_ensemble}
\end{figure}

The final type of uncertainty that will go into our cumulative uncertainty quantification is the uncertainty associated with unmodeled factors. This is referred to as model inadequacy \citep{Kennedy2001} within the discipline of uncertainty quantification. It is also referred to as structural uncertainty as it is connected to the structural stability of a model \citep{Guckenheimer1983}. The outputs from a structurally stable model remain qualitatively similar if the model is perturbed - the perturbations represent low amplitude, unmodeled factors. Structural stability testing can be accomplished using a perturbed physics approach \citep{Astrom2008}. Such an approach can also provide a measure of structural uncertainty for models that do maintain structural stability \citet{Seales2019}. Models with low structural uncertainty can damp perturbations/fluctuations associated with physical factors not directly incorporated into them. Figure \ref{fig:sample_ensemble} shows an example output of such an analysis, henceforth referred to as an ensemble, from a model subjected to a perturbed physics analysis (see \citet{Seales2019} for a full description of this method). For each ensemble, we will use two standard deviations as our uncertainty metric. In performing this analysis, we found that increasing the standard deviation of the perturbation set itself (i.e., the maximum amplitude of perturbations) did not significantly effect the accumulation of uncertainty provided that the perturbations remained randomized in time and of an amplitude below a few percent - an assessment of the uncertainty associated with the particular uncertainty metric itself. Including this as well as all other forms of uncertainty, our analysis involved computing slightly more than 1.25 million model evolutions.

Within our analysis, the success or failure of an ensemble will be determined by comparing the ensemble mean -- the ensemble mean and the unperturbed model solution are equivalent if the model is structurally stable -- and two-standard deviation bounds to paleo and present day constraints. For paleo constraints we use the results of \citet{Ganne2017} who calculated uncertainty bounds on mantle potential temperatures over time (\ref{fig:sample_ensemble}). They derived these bounds by using the MgO content of approximately 22,000 samples of mafic and ultramafic extrusive basalts, and calculating the potential temperature associated with these melts using PRIMELT \citep{Herzberg2015} with different assumed mantle redox conditions. For the present day $T_p$ constraint we use a value of 1350 $^oC$ $\pm 50$ $^oC$ \citep{Herzberg2008}. 

A second present day constraint is the mantle Urey ratio, \emph{Ur}, which is the the ratio of \emph{H} to \emph{Q}. \citet{Jaupart2007} estimate it to be between 0.3 and 0.5. Allowing for continents, the \emph{Ur} upper bound can be extended \citep{Grigne2001, Lenardic2011}. \citet{Lenardic2011} show that for Earth-like continental land fractions heat flows are consistent for mantles with and without continental coverage (for a range that includes Earth's current fraction). The argument is that continental insulation warms the mantle below it and this effect is transmitted to the entire mantle as it is assumed to be well mixed. Increased temperatuers lead to increased plate speed and more efficient cooling of oceanic mantle, offsetting the insulating effect. However, in the continental case, adding the heat producing elements from the continents back into Earth's mantle would increase heat production and therefore \emph{Ur} would also increase by about 0.2. Therefore, the upper \emph{Ur} bound can be extended to approximately 0.7 as the inclusion of a correction to account for continental effects would put this value within data based estimates. We will consider model success with and without continental effects. 

Using the constraints above, we now define the ensemble probability for successful models. This involves identifying the upper and lower most bounds on the ensemble probability distribution that fall within constraints and calculating the probability that an ensemble member falls between these two points. For example, in Figure \ref{fig:sample_ensemble}b the mean of the ensemble is $\sim$1380 $^oC$. The upper temperature bound occurs at 1400 $^oC$, where the present day \emph{Ur} is 0.45, within present day constraints. The lower temperature bound is not set to 1300 $^oC$ because at this temperature \emph{Ur} is greater than 0.5. An \emph{Ur} value of 0.5 occurs at 1365 $^oC$. Therefore, for this ensemble, any output temperature between 1365 and 1400 $^oC$ (the hachured region in \ref{fig:sample_ensemble}b) satisfies present day constraints with a probability of 0.5. This hachured region, then, is the fraction of models within this ensemble that can match present day constraints.  

\begin{table}[h]
\caption{Model Parameters}
\centering
\begin{tabular}{ c c c c }
\hline\hline
Parameter & Values & Units & Description \\
\hline
$T_i$ & 1000, 1250, 1500, 1750, 2000 & $^oC$ & Initial Temperature \\
$T_0$ & 1300, 1350, 1400 & $^oC$ & Scaling Temperature \\
$Q_0$ & 3.0e13, 3.5e13, 4.0e13 & TW & Scaling Heat Flow \\
$H_0$ & 2.19e13, 2.55e13, 2.92e13, 4.38e13, 5.12e13, & TW & Initial Radiogenics \\
& 5.84e13, 6.57e13, 7.66e13, 8.76e13, 1.02e14, & & \\
& 1.09e14, 1.17e14, 1.28e14, 1.46e14 & & \\
$\eta_0$ & 2.21e9 & Pa$\cdot$s & Viscosity constant \\
A & 300 & kJ mol$^{-1}$ & Activation Energy \\
R & 8.314 & J / (mol$\cdot$K) & Universal Gas Constant \\
\hline
\end{tabular}
\label{table:convectiion_parameters}
\end{table}

\section{Results}
\begin{figure}[h!]
    \gridline{\fig{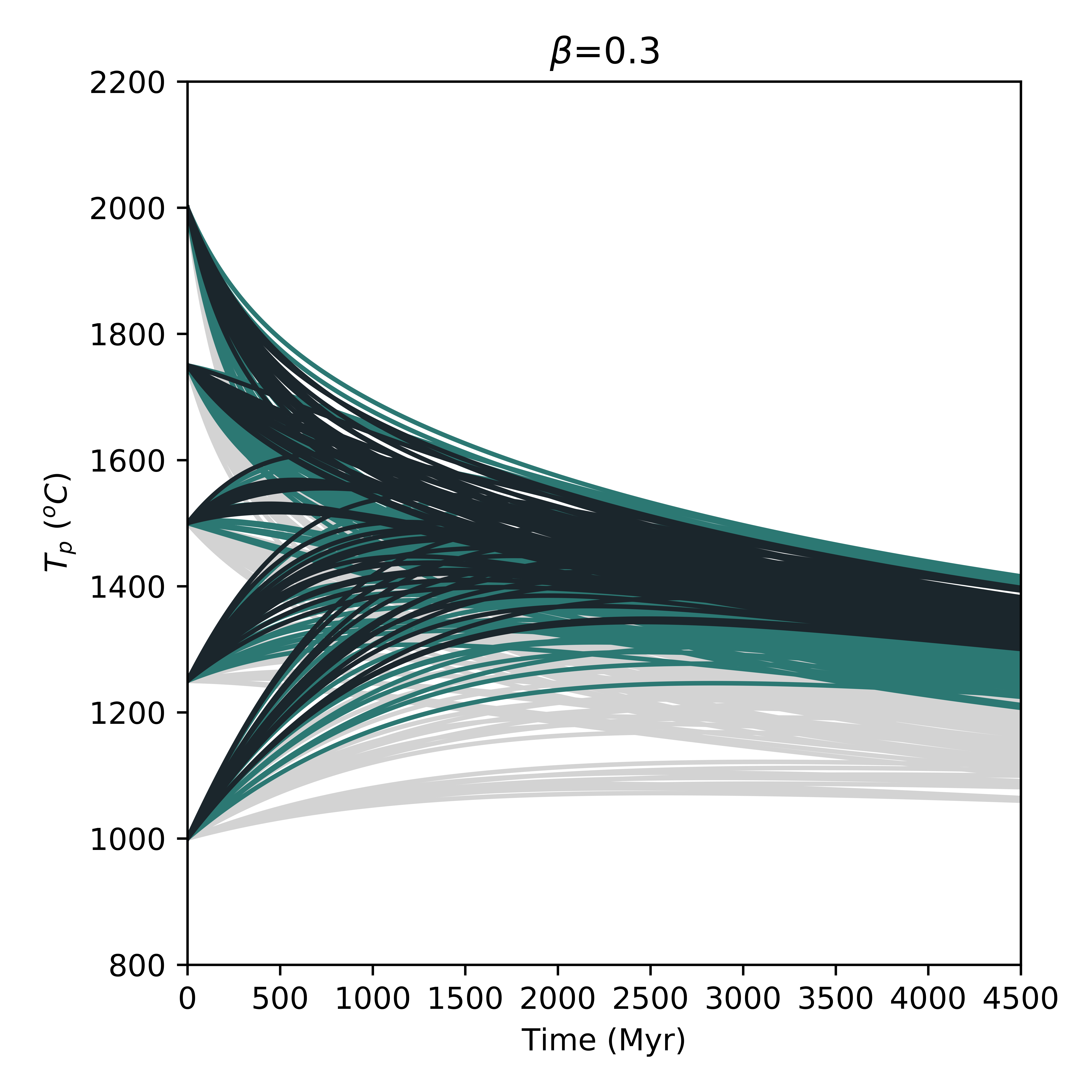}{0.5\textwidth}{(a)}\label{fig:beta_therm_hist_a}
            \fig{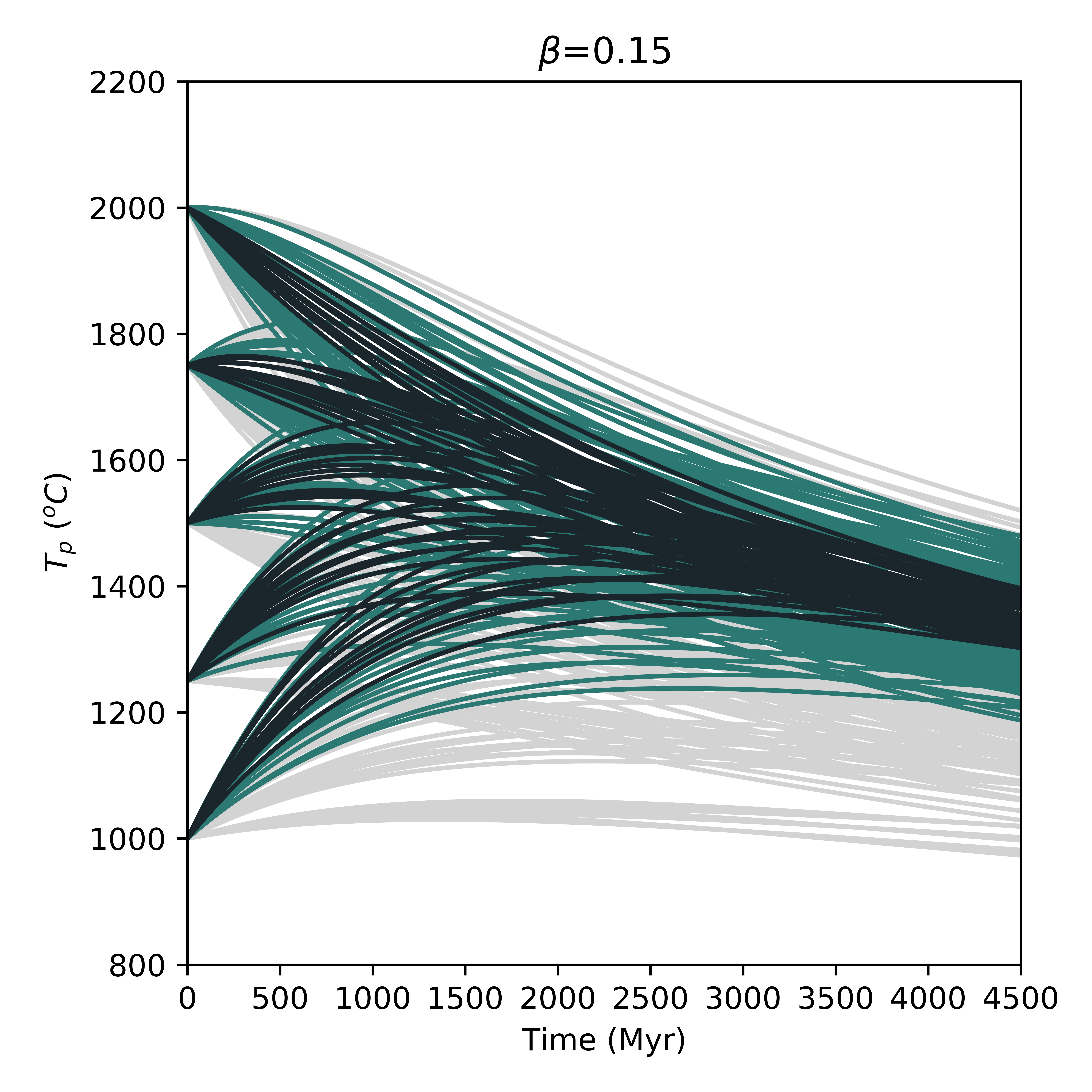}{0.5\textwidth}{(b)}\label{fig:beta_therm_hist_b}}
    \gridline{\fig{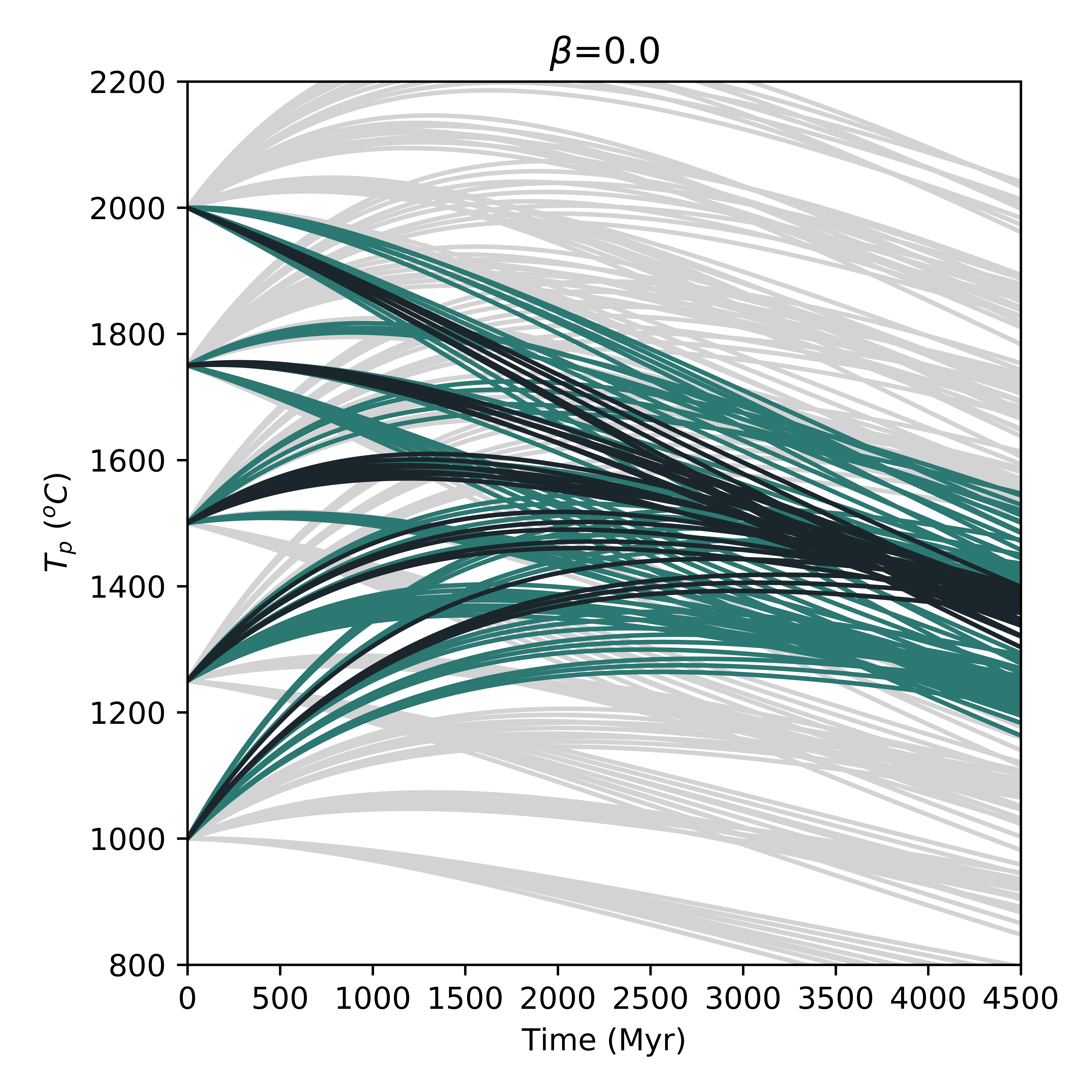}{0.5\textwidth}{(c)}\label{fig:beta_therm_hist_c}
            \fig{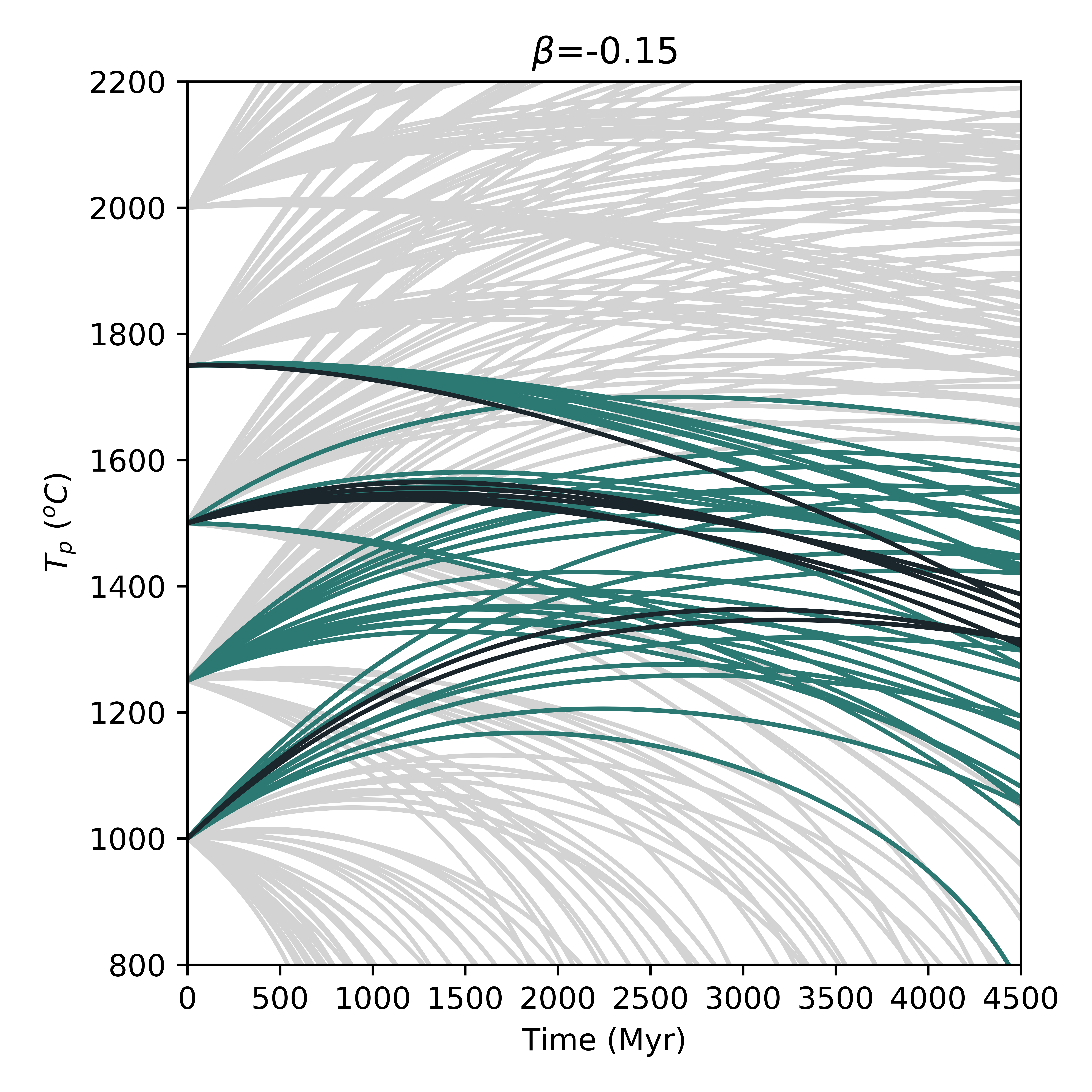}{0.5\textwidth}{(d)}\label{fig:beta_therm_hist_d}}
\caption{Mean ensemble paths from a range of models. Mean ensemble paths are classified into three groups -- those that satisfying an observational constraint (red), those that satisfy it within structural uncertainty (light red), and those that do not satisfying the constraint (gray).}
\label{fig:beta_therm_hist}
\end{figure}

Figure \ref{fig:beta_therm_hist} shows mean ensemble cooling paths for each tested ensemble for models with different $\beta$ values, input parameter values, and initial conditions. We leave off the ensemble uncertainty bounds for clarity and ease of viewing, but they were calculated for all the plotted ensembles. The mean ensemble paths that satisfy the present day $T_p$ constraints are shown as red lines. Mean ensemble paths that fall outside the constraint, but are associated with models that can match the constraint within structural uncertainty bounds, are shown as light red lines. Solutions that do not match the constraint, even allowing for ensemble uncertainty bounds, are shown as grey lines. A model with $\beta<0$ is very initial condition and input value dependent. This leads to a wide model solution space. Models with $\beta\geq 0$ had weaker initial condition and input value dependencies, resulting in a more concentrated solution space. 

\begin{figure}[h!]
  \centering
  \includegraphics[width=0.5\linewidth]{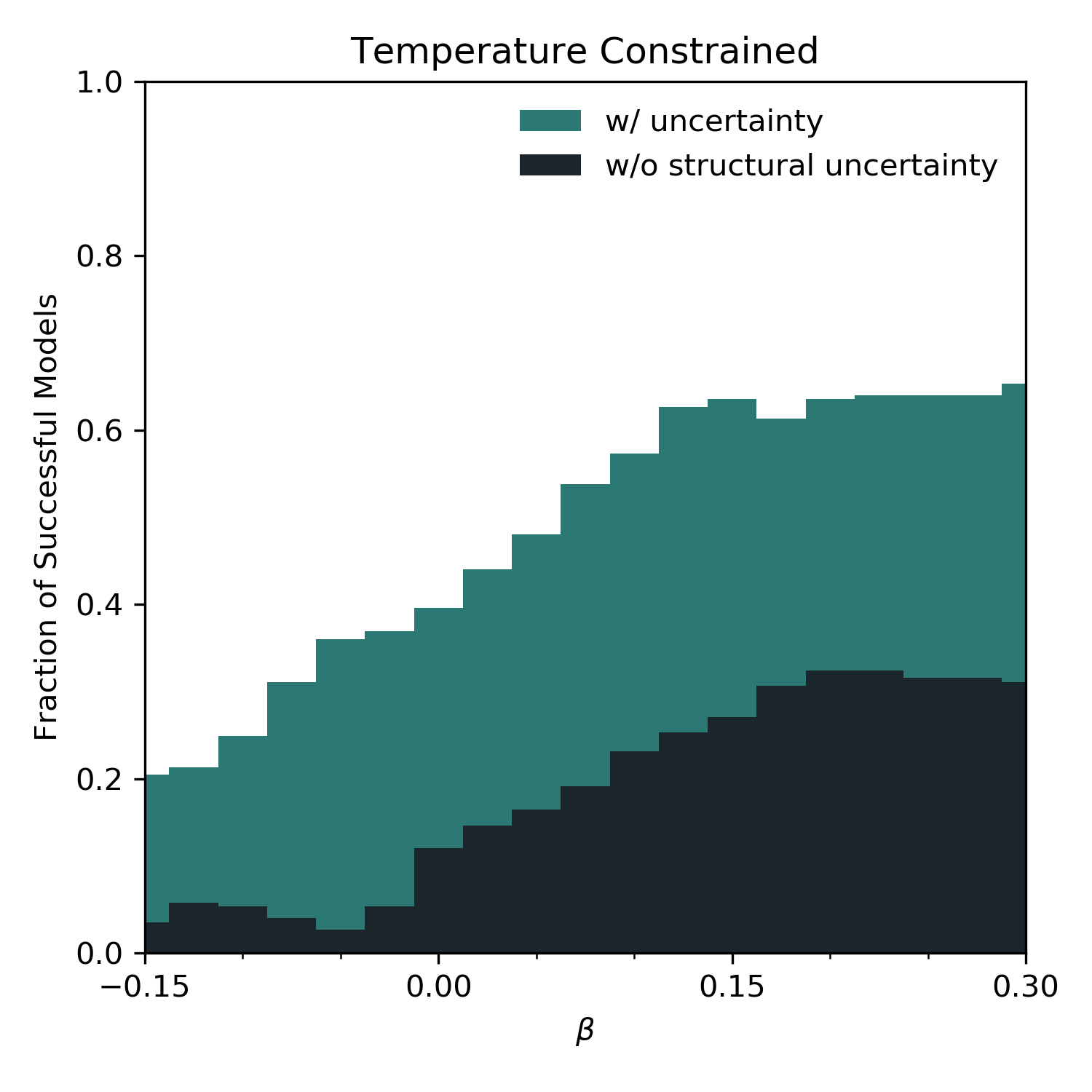}
  \caption{Probability distribution for models that satisfy the present day temperature constraint.}
  \label{fig:beta_presentT}
\end{figure}

The number of cases that satisfy the present day $T_p$ constraint for variable $\beta$ are shown in Figure \ref{fig:beta_presentT}.  The number of cases where the mean matches the present day constraint (darker green) is small at the most negative $\beta$ endmember. The number of mean ensemble paths remained below 10\% until $\beta$ became positive and the number of successful cases began to grow. At a $\beta$ value near 0.2 the number of successful cases plateaued around 30\%. Accounting for structural uncertainty increases the number of successful cases for all $\beta$ values (lighter green). The rise in successful cases occurred while $\beta$ was still negative, around $\beta=-0.1$, and plateaued at a $\beta$ slightly greater than 0.1.

\begin{figure}[h!]
  \centering
  \includegraphics[width=0.5\linewidth]{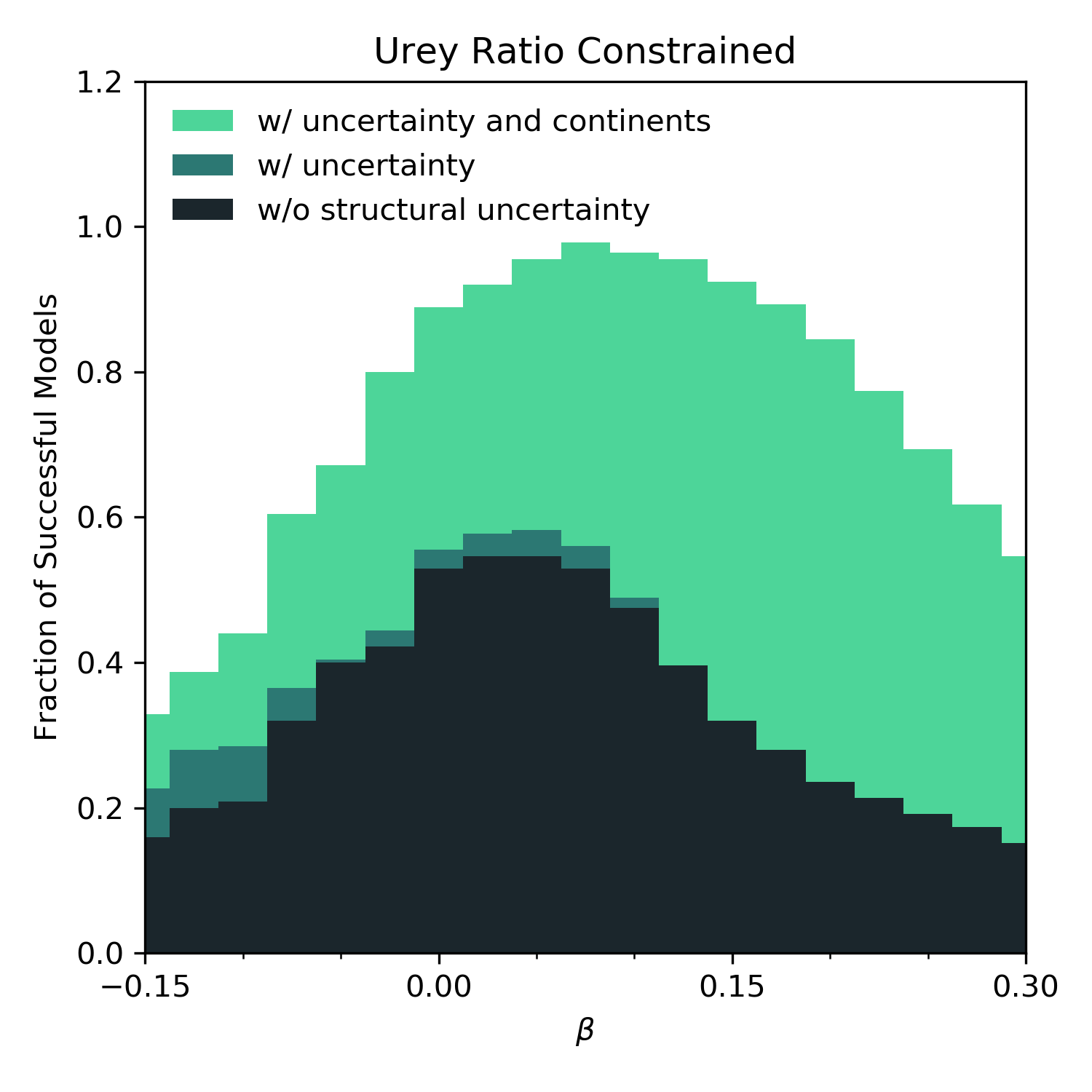}
  \caption{Probability distribution for models that satisfy the present day Urey ratio constraint.}
  \label{fig:beta_presentUr}
\end{figure}

The number of cases matching present day \emph{Ur} are shown in Figure \ref{fig:beta_presentUr}. The color scheme is the same as Figure \ref{fig:beta_presentT} with mean solutions in darker green and those that include structural uncertainty in lighter green. Results are shown for cases that match present day \emph{Ur} without accounting for the effect of present day continental distribution (lighter green) and for cases in which the effect of continents, on the present day Urey ratio \citep{Grigne2001,Lenardic2011}, is accounted for (lightest green). The distribution of successful cases peaked around 60\% for $\beta=0.05$. Accounting for structural uncertainty had little effect for the \emph{Ur} constraint. Accounting for continents increased the number of successful cases. At its peak, near $\beta=0.1$, the number of successful matches was greater than 90\%. Including a continental effect  disproportionately benefited more positive $\beta$ values. 

\begin{figure}[h!]
  \centering
  \includegraphics[width=0.5\linewidth]{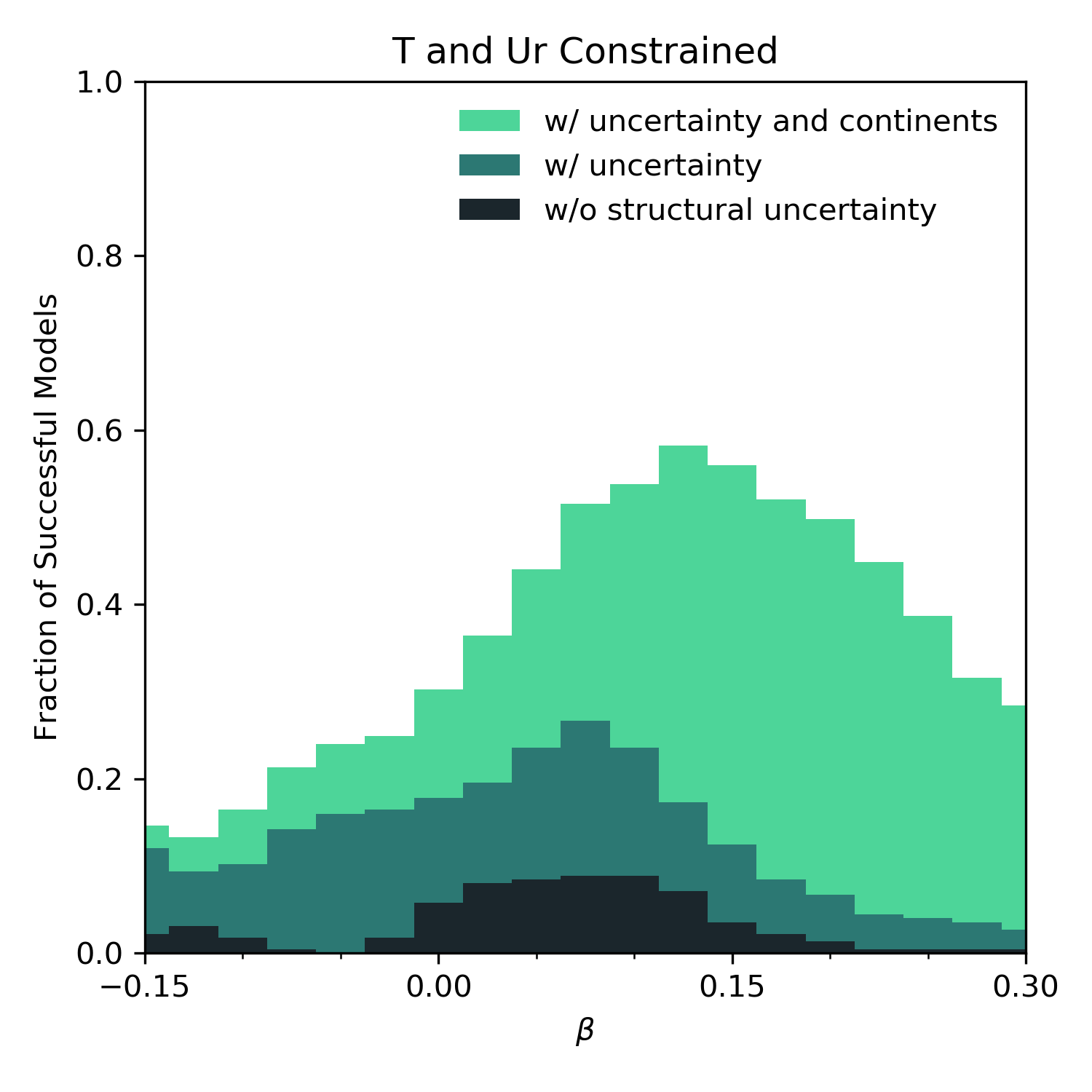}
  \caption{Probability distribution for models that satisfy the present day temperature and Urey ratio constraint.}
  \label{fig:beta_presentTUr}
\end{figure}

Figure \ref{fig:beta_presentTUr} shows the number of successful cases when assigning equal weight to the present day $T_p$ and \emph{Ur} constraints. The distribution is non-normal. Mean ensemble paths resulted in less than 10\% of successful cases across the board. The peak for the mean solutions is at a $\beta$ value slightly less than 0.1. Below this value successful models fall to nearly zero before increasing slightly when values of $\beta<0$ were considered. Increasing the upper \emph{Ur} bound to 0.7, to account for the potential effect of continents, shifted the peak $\beta$ value to be greater than 0.1 and increased the number of successful cases to nearly 60\% at the peak. Considering structural uncertainty preferentially benefited the lower half of the tested $\beta$ space. A very low percentage of models could match both constraints for $\beta$ values greater than 0.2 unless the effects of continents were considered (and it should be kept in mind that doing so adds its own layer of uncertainty as the continental correction comes from models \citep{Grigne2001,Lenardic2011}).

\begin{figure}[h!]
  \centering
  \includegraphics[width=0.5\linewidth]{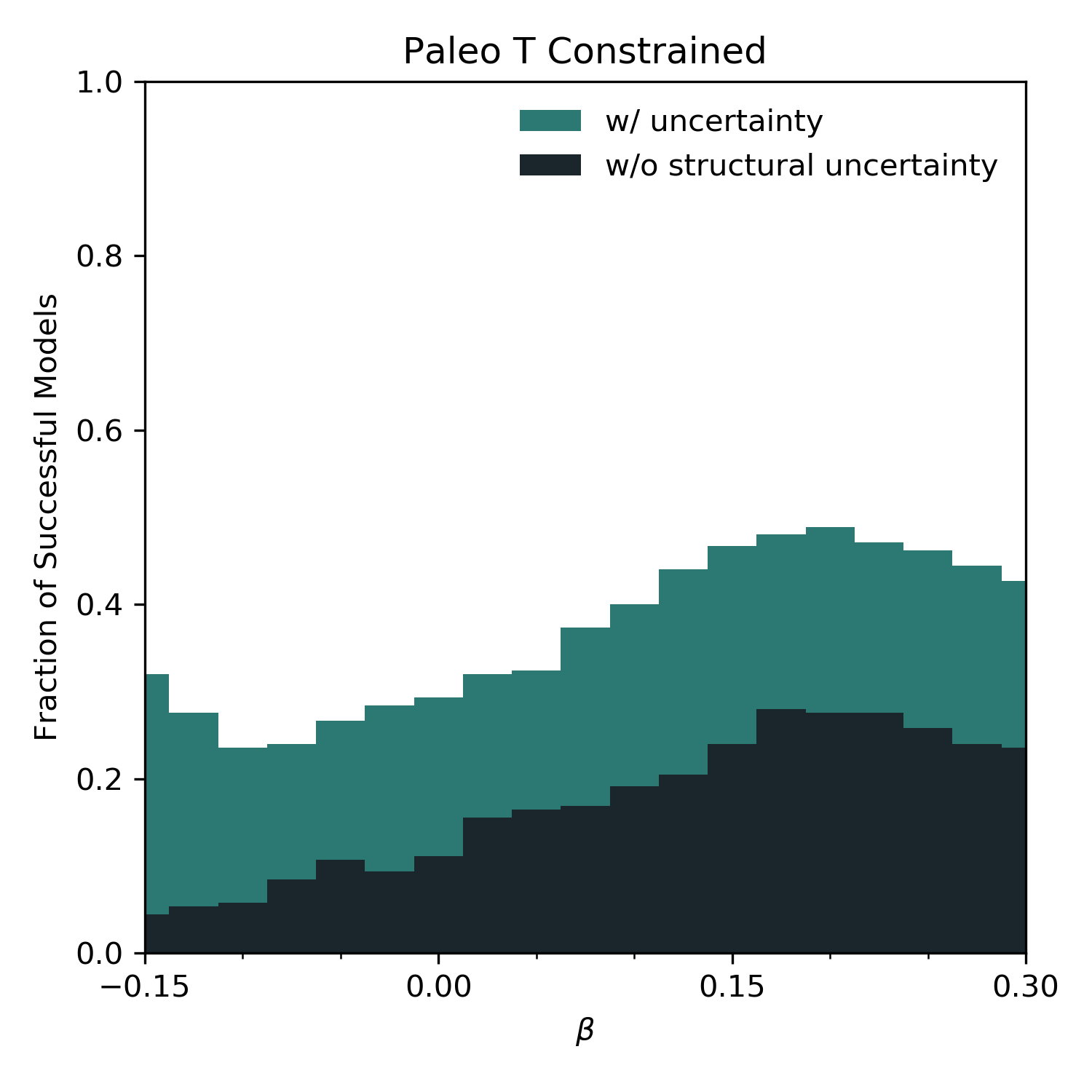}
  \caption{Probability distribution for models that satisfy the paleo temperature constraints.}
  \label{fig:beta_paleo}
\end{figure}

The distributions that resulted from using only paleo temperature constraints are shown in Figure \ref{fig:beta_paleo}. The trends are similar to those in Figure \ref{fig:beta_presentT}. One difference is the uniform decrease in the fraction of ensembles able to match the paleo constraints. This intuitively makes sense in that to be successful an ensemble must stay within a temperature window over an extended evolution time rather than match a value at a single time. A subtle, but noteworthy difference between Figure \ref{fig:beta_presentT} and Figure \ref{fig:beta_paleo} is that the fraction of successful ensembles matching the paleo constraints increased to a greater degree as $\beta$ was increased. Positive $\beta$ models tend to lessen initial condition dependence. Nearer to the model start time there is less time to eliminate the influence of the initial condition. As a result, some models that converge to present day temperatures were too hot or too cold at 2.5 Gyr and thus considered unsuccessful. Even with this change in slope, the distribution peaked around 0.2.

\begin{figure}[h!]
    \gridline{\fig{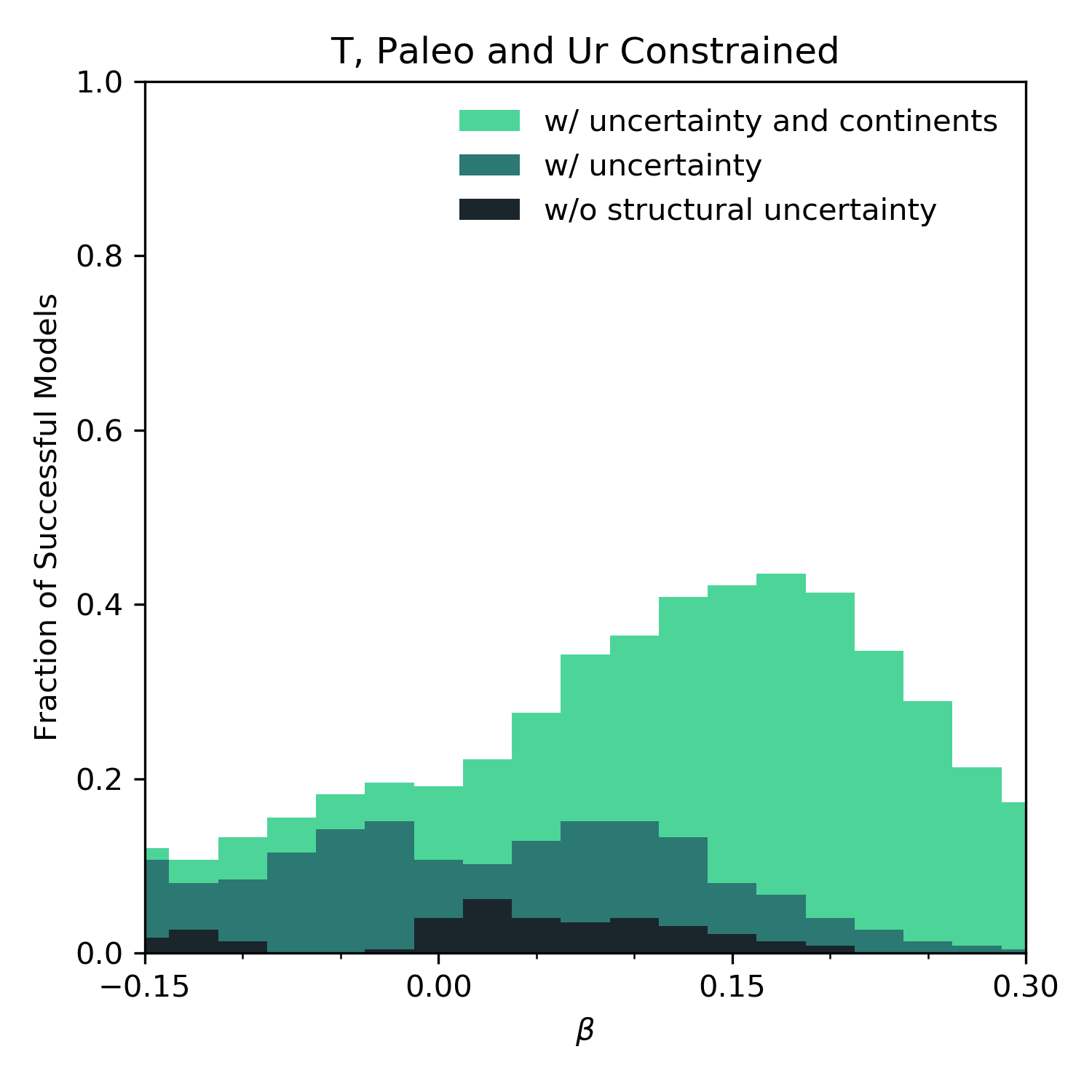}{0.5\textwidth}{(a)}\label{fig:allconsta}
            \fig{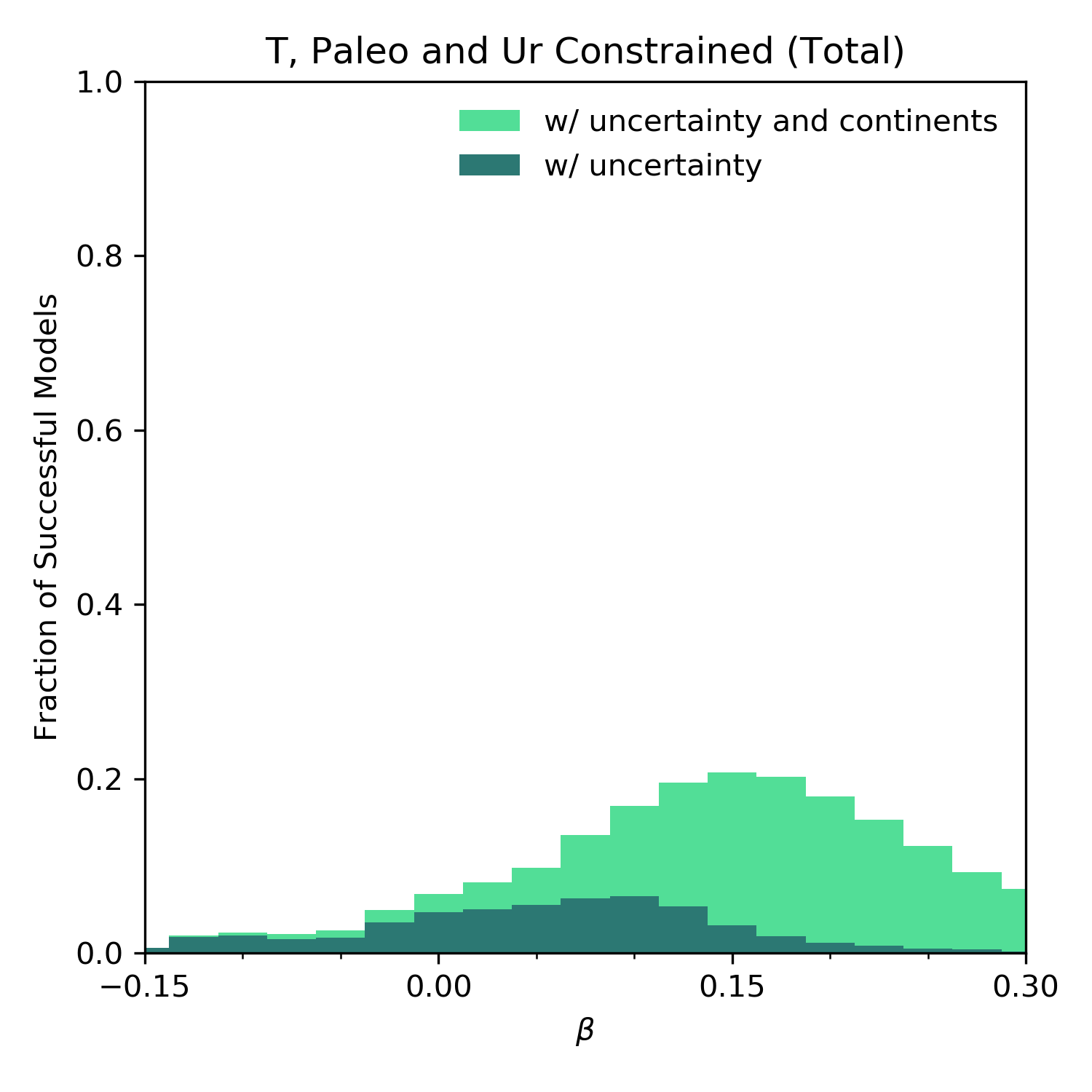}{0.5\textwidth}{(b)}\label{fig:allconstb}}
\caption{Probability distribution for models that satisfy all observational constraints (a). In (b we tabulate the total fraction of models that match all constraints by calculating the probability of success of each ensemble as shown in Figure \ref{fig:sample_ensemble}} 
\label{fig:all_const}
\end{figure}

Figure \ref{fig:all_const} shows models that can match paleo and present day constraints. Figure \ref{fig:allconsta} shows the fraction of models for each $\beta$ that have some portion of the ensemble that satisfies all three constraints. Distributions are bi-modal, having one peak in the negative $\beta$ domain and one peak in the positive domain. Accounting for structural uncertainty increased the fraction of successful solutions across the board and produced nearly identical peaks in both the positive and negative domains. Allowing for continental effects shifted the largest peak close to a $\beta$ value of 0.2, but a peak just less than zero remained. A representation of the total probability is shown in Figure \ref{fig:allconstb}. For each $\beta$, the total probability is the sum of each ensemble probability divided by the total number of initial conditions and model input combinations assessed. For a constraint on present day \emph{Ur} that does not account for continents, a peak probability of approximately 10\% occurred at a $\beta$ value of 0.1. This distribution has a single peak with a heavy left tail, which is caused by the hard upper \emph{Ur} limit of 0.5 that cast out a large portion of the more positive $\beta$ ensembles. Relaxing this constraint resulted in more normal distribution peaked around 0.15. This is close to the value argued for by \citet{Conrad1999}. We have given all data constraints equal weight. If one of the constraints is found to be more reliable than the others, then the distribution peak will shift towards the $\beta$ values that coincide with matching that constraint. 

\begin{figure}[h!]
    \gridline{\fig{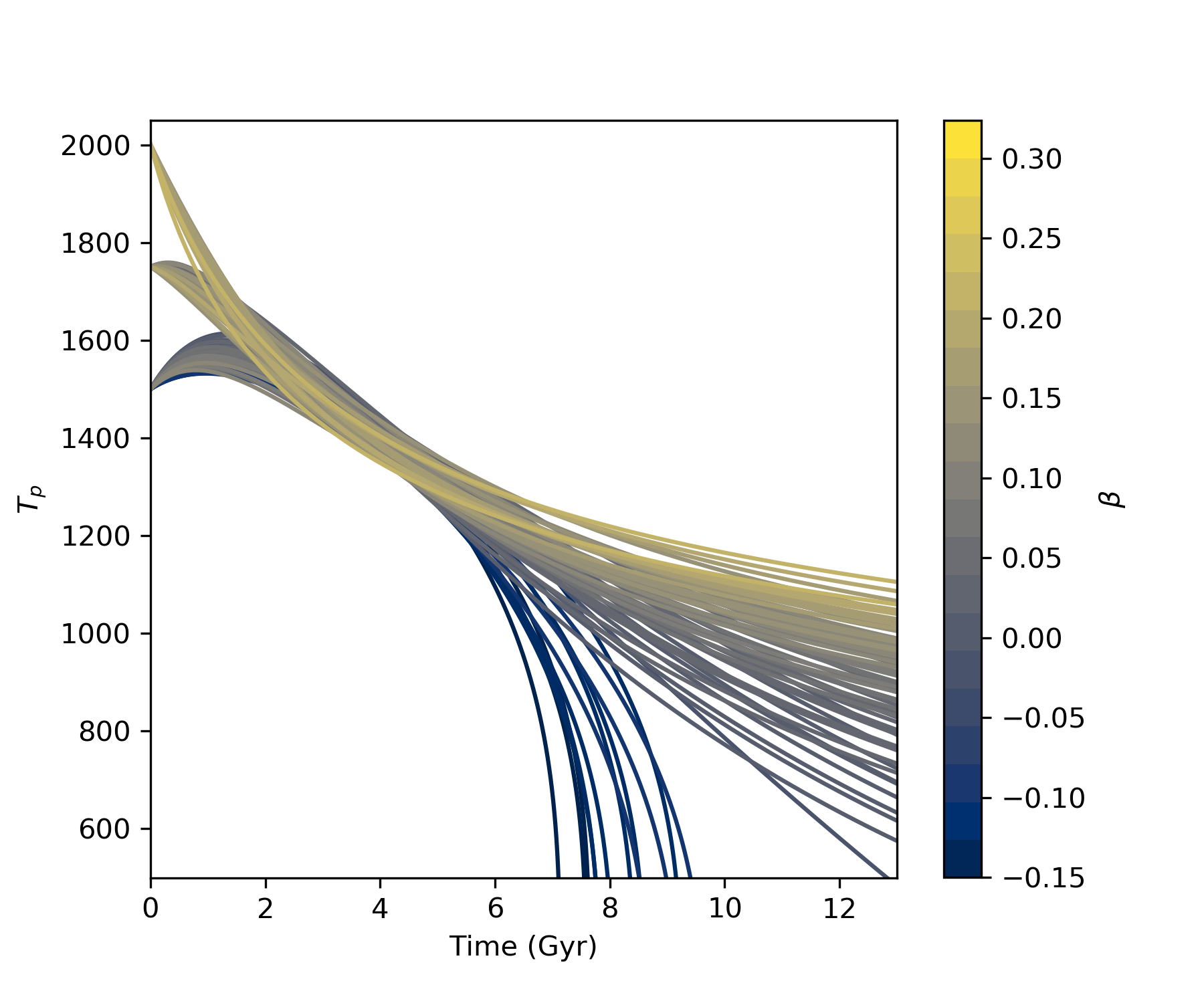}{0.5\textwidth}{(a)}\label{fig:beta_proja}
            \fig{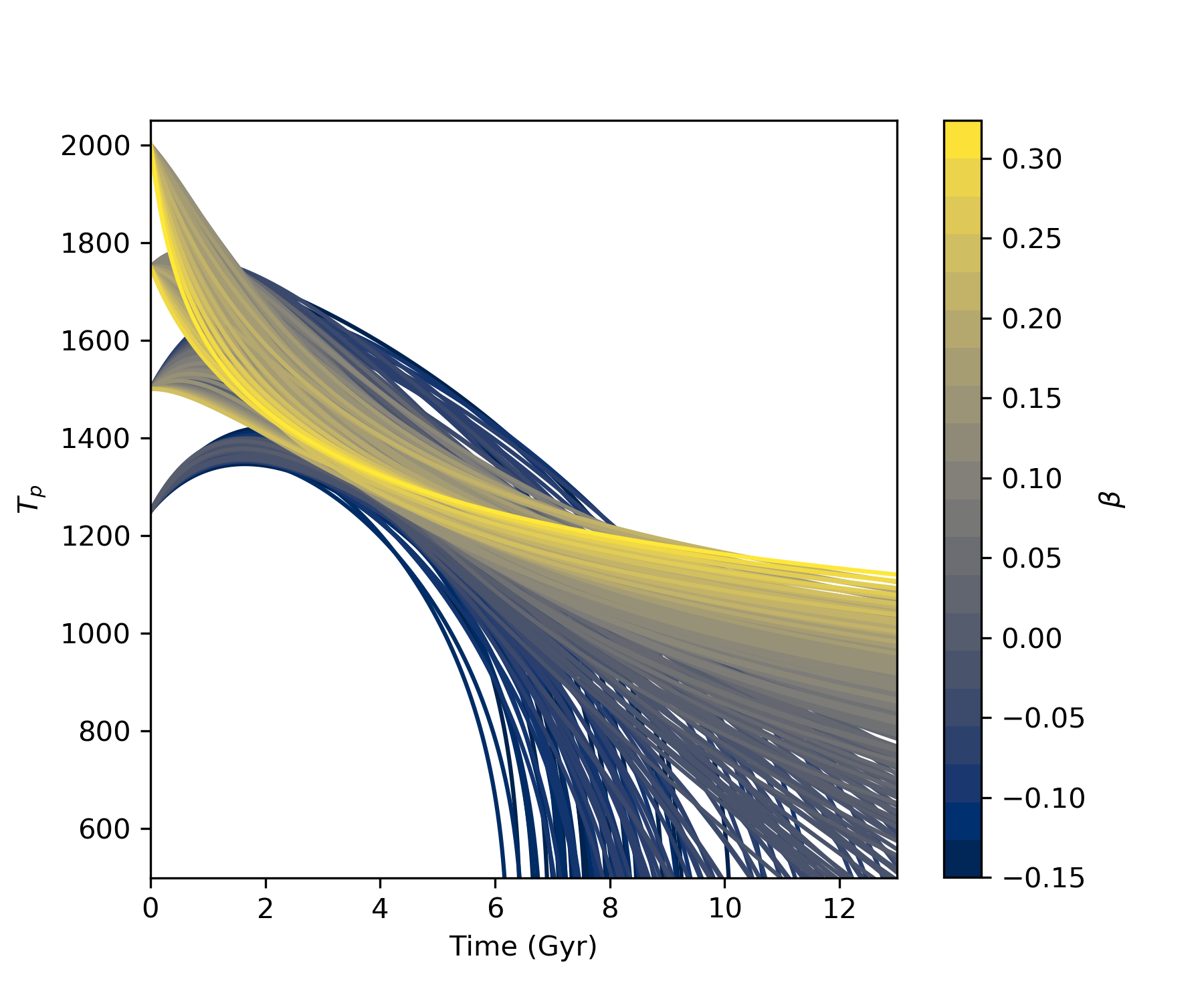}{0.5\textwidth}{(b)}\label{fig:beta_projb}}
\caption{Forward in time projections for successful models. In (a) only mean ensemble paths that were successful without considering ensemble uncertainty are plotted (i.e., model structural uncertainty is not accounted for). In (b) mean ensemble paths that were successful within ensemble uncertainty are plotted (i.e., this accounts for structural uncertainty).} 
\label{fig:beta_proj}
\end{figure}

Using only the mean ensemble paths that matched paleo and present day constraints, we projected mantle potential temperature out to 10 Gyr (Figure \ref{fig:beta_proj}). Figure \ref{fig:beta_proj}a projects only those mean ensemble paths that matched Earth constraints (darker green models in Figure \ref{fig:allconsta}). The differing feedbacks within the models become apparent as time evolves with negative $\beta$ models (positive feedback) reaching far cooler mantle temperatures. These models lead to cooling runaways and once temperatures drop too low the models are cut off as they have lost structural stability, that is small perturbations/fluctuations significantly affect model evolution, pulling the perturbed solution far from the unperturbed solution \citep{Seales2019}. Models with $\beta>0.1$ cooled more slowly, maintaining temperatures above 1000 $^oC$ throughout. Figure \ref{fig:beta_proj}b shows the projected models that match paleo and present day constraints with structural uncertainty now accounted for. Projections were limited to those models that matched \emph{Ur} values between 0.3 and 0.5. Including structural uncertainties allowed for run away cooling behavior to occur nearly one billion years nearer to present day for models with the most negative $\beta$ values. If we take into account the total probabilities, which peak between $\beta$ values of 0.1 and 0.2, and only use those cases, then projected temperature vary between 1000 and 1200 $^o$ C at 10 Gyr of model evolution. However, as each of the mean ensemble paths plotted match Earth constraints, they all remain possible. Stated another way, there is no reason why the evolution path of a particular planet, the Earth, needs to follow a most probable path within a model solutions space. 

\section{Discussion}
Our analysis considered multiple forms of uncertainty to assess the probability that any given model fits Earth constraints. Any model with a probability greater than zero is capable of explaining Earth's thermal history. One model being less probable than another, in model solution space, does not eliminate the possibility that the lower probability model captures the essential physics of plate tectonics, as related to planetary cooling. Having said that, we can also weigh probabilities to assess which models can match Earth constraints over the widest range of uncertainties. Figure \ref{fig:allconstb} indicates models with $\beta$ between 0.1 and 0.2 fall into this category. High $\beta$ models can match present day temperature over a wide input range but struggle to match the lower \emph{Ur} constraint. Lower $\beta$ models can match the \emph{Ur} constraint but struggle with present day temperature constraints if $\beta$ drops too low as they then run hot \citep{McNamara2000}. That a ``sweet spot" could exist between the two end-members is not, in hindsight, qualitatively surprising. 

Uncertainties in the data constraints we used influenced the calculated probability of successful models. We assumed equal weighting for each constraint. Of the two present day constraints, the present day mantle temperature is a harder constraint. This is because there is less uncertainty in estimating its value than there is in estimating the Urey ratio \citet{Jaupart2007}. The difficulty of considering different weightings is that, although the distribution of uncertainties associated with temperature data has been calculated \citep{Condie2016,Ganne2017}, the same is not true for the Urey ratio. At this stage, we did not consider it warranted to apply different weightings but this could be done in the future.

Our analysis explored a slice of potential model space. A more extensive exploration would change quantitative results but key qualitative results are likely to be robust. The qualitative differences between positive and negative $\beta$ models comes from the fact that the former is dominated by a negative system feedback and the latter by a positive feedback \citep{Moore2015}. More sophisticated models can be constructed (e.g., fully 3-D models) but the dominant feedback will still dictate end-member behavior and uncertainty structure. Uncertainties will be smaller in models dominated by negative feedbacks, as will be model solution space. The latter means that model cases may be less likely to match any given data constraints, but if they can match constraints, then a narrow solution space will lead to a larger percentage of model ensembles being successful. Models with high positive feedbacks will have greater uncertainty and an associated larger solution space. A large solution space increases the potential that at least some cases can match a given data constraint and, at the same time, it favors a smaller percentage of potential model ensembles being successful. 

The connection between uncertainty and successful models relates to another conclusion we argue is robust: Models based on different hypotheses, regarding the dynamics of plate tectonics, are consistent with constraints on the Earth's thermal history, i.e., competing hypotheses remain viable. Phrased another way, model and data uncertainties lead to ambiguity - more than one model is viable. Considering more sophisticated (complex) models will not, we argue, change this conclusion, provided full model uncertainties are assessed. Increasing complexity can increase model uncertainty \citep{Saltelli2019}. More complex models come with more parameters and assumptions which increases both parametric and model selection uncertainty \citep{Saltelli2019}, as well as the potential to overfit data. That can increase the number of potential model solutions and the computational time needed to find them. It can also greatly increase the time and work load needed to quantify model uncertainty. More complex models may be able to better match data constraints but this should not be confused with the models being more certain. The ability of a model to match constraints is not the same thing as a model's uncertainty. Model uncertainty can, however, affect the ability of a model to match constraints. More uncertain models are associated with a larger potential model solution space. A larger solution space increases the potential that some combinations of model inputs, initial conditions, and ensemble paths will match constraints. It is possible that new and/or more certain data constraints could bridle this to a degree, though historical data from the Earth will always have uncertainty. As such, we argue that multiple hypotheses will likely remain viable into the near future, particularly if there is a trend toward developing more complex models.  

Model ambiguity in Earth science is not new \citep[e.g.,][]{Richards2018} and it has been considered for endmember cases in thermal history modeling \citep{Korenaga2008}. Accounting for model uncertainties extends the range of model ambiguity such that multiple hypotheses, regarding the dynamics of planetary cooling, can be consistent with data constraints. Hypothesis discrimination can continue, but it must proceed in a statistical manner. This, we argue, is another robust conclusion. We can ask which models come with higher probabilities of success in light of uncertainties. This is the utility of Figure \ref{fig:all_const}. The degree to which one is willing to push this further depends on a question that cannot be scientifically answered at present: of all the possible evolution paths, consistent with physical and chemical principles, did a single planet, the Earth, follow what is the most likely path in that potentiality space? The conservative stance is to say 'We don't know,' which means we consider all models with greater than zero probability as viable. 

The question above relates to the extension of thermal history studies from Earth to planetary application, habitability in particular \citep{Kite2009,Schaefer2015,Komacek2016,Foley2015,Foley2016,Tosi2017,Foley2017,Barnes2020}. Thermal history models applied to the Earth are postdictive: they set out to match historical data. In the context of habitability studies, thermal history models are used in a predictive mode to determine whether liquid water may be present on the surface of terrestrial planets with variable planetary and orbital properties. These predictions are made by calculating the flux of volatiles from the interior of the planet to the surface using mantle temperatures along with melting modules and testing whether surface water can persist over time scales that allow life to develop. Using models in a predictive mode increases the potentiality space of model outputs. With the thought of limiting the vastness of this space, many studies have focused on planets similar to Earth in size and composition as a starting point \citep[e.g.,][]{Foley2015,Foley2016,Rushby2018}. Implicit to this is the thought that uncertainties will be lowest for modeling this subset of planets. Our analysis suggests that even if we consider the most Earth-like planet possible, with the most observational data (the Earth), significant uncertainty remains [Fig. 9]. 

The above leads to a few suggestions on moving forward. First, even if we focus on a plate tectonic mode of planetary cooling, we should consider all viable models [Figure \ref{fig:all_const}]. To date, habitability models have considered a single plate tectonic model \citep[e.g.,][]{Driscoll2013,Foley2015,Foley2016,Rushby2018}. The particular model adopted (a classic high $\beta$ model) is not the most probable model for matching Earth data. This is not damning but it is inconsistent with the idea that many of the studies are based on: given a large model space, let's start with models that best account for Earth data. It also bypasses model selection uncertainty. Second, all models should be subjected to a more robust uncertainty analysis. Typically only a range of initial conditions and input values are tested. An ensemble approach is generally not employed, which leaves out structural uncertainty. One uncertainty measure is not a substitute for another and all need to be evaluated before model implications can be assessed and/or before a model can be validated. A corollary is that model implications need to be viewed in a probabilistic manner by presenting results as probability distributions. This becomes particularly important for models used to make forecasts. 

All of the projections in Figure \ref{fig:beta_proj}b should, we argue, be considered as potentialities. In that view, they are all counterfactuals \citep{Taleb2012} with very different implications if used as forecasts. For example, a family of ensembles imply that plate tectonics could end in about 1.5 billion years as the mantle becomes too cold, transitioning out of plate tectonics and eventually shutting down melting. This family of ensembles is consistent with a study that did forecast the end of plate tectonics in 1.45 billion years \citep{Cheng2018}. Such a forecast has implications for life beyond Earth. The fact that some of our projections are in line with the study of \citet{Cheng2018} speaks to model reproducibility, as that study used a negative $\beta$ model, which is also the one we found leads to cold runaways. However, it is the negative $\beta$ models that are associated with the largest uncertainty and are prone to structural instability \citet{Seales2019}. Not being clear about uncertainty, especially for a provocative conclusion, only invites misinformation (e.g., presenting a highly uncertain model forecast as a singular "result"). We would suggest that if full uncertainty analysis was as strong a component of planetary modeling studies as, for example, a methods section, then the odds of unintentionally making conclusions that can send misinformation would be reduced. We will add a corollary, the greater a modeling study moves toward the prediction end of the postdiction/description-prediction/forecast spectrum, the greater the responsibility of the modelers to present a full uncertainty analysis. That corollary applies to essentially all modeling studies of terrestrial exoplanets. Adhering to it could, we argue, prevent red-herring debates of the type that have surfaced in the past \citep[e.g.,][]{Chorost2013}.

In the exoplanet modeling field, thermal history models are being coupled to other models to explore how interior planet evolution co-evolves with other systems -- stellar, orbital, volatile cycling, climate, weathering and life \citep{Barnes2020}. Each system sub-model is subject to the types of uncertainty we have presented for thermal history models, making the full model potentiality space large. This can make a grid search approach, to map out the coupled model solution space in light of uncertainties, intractable. However, the full model potentiality space is often not of primary interest. A more primary driver behind the coupled models is mapping the subspace that allows water to exist at the surface of a planet over geologic time (this connects the models to the search for life beyond Earth – life as we know it relies on water). Having a search target, within model potentiality space, can reduce the computational work load, but a grid search, akin to that of this paper, would still be impractical given the large dimensionality of the problem. More efficient computational methods can bring the modeling back to a tractable level (e.g., machine learning based methods \citep{Fleming}). This will introduce further uncertainty that will need to be evaluated -- the uncertainty associated with the particular search method. All of this will increase the workload and the move toward a probabilistic framework. Such a framework, in turn, would move the field beyond a binary assessment habitability and towards assessing the potential of a planet to host life that requires a particular type of environment. Given that all of this is being done in the prediction/forecast mode, uncertainty analysis will need to play a larger role than it has to date. 

\section{Conclusions}
We applied an uncertainty analysis to solid Earth cooling models. The analysis accounted for the combined effects of: 1) Model selection uncertainty; 2) Model structural uncertainty; 3) Uncertainty in initial conditions; 4) Uncertainty in model input values. Accounting for model and observational uncertainties allows for model validation (testing the degree to which model outputs can match data constraints). Validation, once full uncertainty measures are evaluated, requires a probabilistic approach and results are presented as probability distributions. Given we only have one planet evolutionary path, the Earth, we have argued that any models that maintain finite probabilities of accounting for observational data, over model potentiality space and in light of uncertainties, remain viable. For the thermal history models we examined this leads to ambiguity (multiple hypotheses remain viable for the Earth's thermal history). When  thermal history models move from a postdictive mode (accounting for existing Earth data) into a predictive mode designed to constrain conditions that allow for clement surface environments on terrestrial planets, the uncertainty analysis becomes more critical. 

\bibliography{Refs2.bib}

\begin{thebibliography}{}
\expandafter\ifx\csname natexlab\endcsname\relax\def\natexlab#1{#1}\fi

\bibitem[{Astrom \& Murray(2008)}]{Astrom2008}
Astrom, K. J. K.~J., \& Murray, R.~M. 2008, {Feedback systems: an introduction
  for scientists and engineers} (Princeton University Press), 408,
  doi:10.5860/choice.46-2107

\bibitem[{Barnes {et~al.}(2020)Barnes, Luger, Deitrick, Driscoll, Quinn,
  Fleming, Smotherman, McDonald, Wilhelm, Garcia, Barth, Guyer, Meadows, Bitz,
  Gupta, Domagal-Goldman, \& Armstrong}]{Barnes2020}
Barnes, R., Luger, R., Deitrick, R., {et~al.} 2020, Publications of the
  Astronomical Society of the Pacific, 132, 024502

\bibitem[{Baross \& Hoffman(1985)}]{Baross1985}
Baross, J.~A., \& Hoffman, S.~E. 1985, Origins of Life and Evolution of the
  Biosphere, 15, 327

\bibitem[{Cheng(2018)}]{Cheng2018}
Cheng, Q. 2018, Gondwana Research, 63, 268

\bibitem[{Chorost(2013)}]{Chorost2013}
Chorost, M. 2013, Astronomy Now, 18

\bibitem[{Christensen(1985)}]{Christensen1985}
Christensen, U.~R. 1985, Journal of Geophysical Research, 90, 2995

\bibitem[{Condie {et~al.}(2016)Condie, Aster, \& {Van Hunen}}]{Condie2016}
Condie, K.~C., Aster, R.~C., \& {Van Hunen}, J. 2016, Geoscience Frontiers, 7,
  543

\bibitem[{Conrad \& Hager(1999{\natexlab{a}})}]{Conrad1999b}
Conrad, C.~P., \& Hager, B.~H. 1999{\natexlab{a}}, Journal of Geophysical
  Research: Solid Earth, 104, 17551

\bibitem[{Conrad \& Hager(1999{\natexlab{b}})}]{Conrad1999}
---. 1999{\natexlab{b}}, Geophysical Research Letters, 26, 3041

\bibitem[{Curry \& Webster(2011)}]{Curry2011}
Curry, J.~A., \& Webster, P.~J. 2011, {Climate science and the uncertainty
  monster},  American Meteorological Society, doi:10.1175/2011BAMS3139.1

\bibitem[{Davies(1980)}]{Davies1980}
Davies, G.~F. 1980, Journal of Geophysical Research, 85, 2517

\bibitem[{Driscoll \& Bercovici(2013)}]{Driscoll2013}
Driscoll, P., \& Bercovici, D. 2013, Icarus, 226, 1447

\bibitem[{Feulner(2012)}]{Feulner2012}
Feulner, G. 2012, Reviews of Geophysics, 50, doi:10.1029/2011RG000375

\bibitem[{Foley(2015)}]{Foley2015}
Foley, B.~J. 2015, Astrophysical Journal, 812, 36

\bibitem[{Foley \& Driscoll(2016)}]{Foley2016}
Foley, B.~J., \& Driscoll, P.~E. 2016, {Whole planet coupling between climate,
  mantle, and core: Implications for rocky planet evolution},  John Wiley {\&}
  Sons, Ltd, arXiv:1711.06801

\bibitem[{Foley \& Smye(2018)}]{Foley2017}
Foley, B.~J., \& Smye, A.~J. 2018, Astrobiology, 18, 873

\bibitem[{Ganne \& Feng(2017)}]{Ganne2017}
Ganne, J., \& Feng, X. 2017, Geochemistry, Geophysics, Geosystems, 18, 872

\bibitem[{Giannandrea \& Christensen(1993)}]{Giannandrea1993}
Giannandrea, E., \& Christensen, U. 1993, Physics of the Earth and Planetary
  Interiors, 78, 139

\bibitem[{Grign{\'{e}} \& Labrosse(2001)}]{Grigne2001}
Grign{\'{e}}, C., \& Labrosse, S. 2001, Geophysical Research Letters, 28, 2707

\bibitem[{Guckenheimer \& Holmes(1983)}]{Guckenheimer1983}
Guckenheimer, J., \& Holmes, P.~J. 1983, {Nonlinear Oscillations, Dynamical
  Systems, and Bifurcations of Vector Fields}, Applied Mathematical Sciences
  (New York, NY: Springer New York), 462, doi:10.1007/978-1-4612-1140-2

\bibitem[{Gurnis(1989)}]{Gurnis1989}
Gurnis, M. 1989, Geophysical Research Letters, 16, 179

\bibitem[{Herzberg \& Asimow(2008)}]{Herzberg2008}
Herzberg, C., \& Asimow, P.~D. 2008, Geochemistry, Geophysics, Geosystems, 9,
  n/a

\bibitem[{Herzberg \& Asimow(2015)}]{Herzberg2015}
---. 2015, Geochemistry, Geophysics, Geosystems, 16, 563

\bibitem[{H{\"{o}}ink {et~al.}(2011)H{\"{o}}ink, Jellinek, \&
  Lenardic}]{Hoink2011}
H{\"{o}}ink, T., Jellinek, A.~M., \& Lenardic, A. 2011, Geochemistry,
  Geophysics, Geosystems, 12, n/a

\bibitem[{H{\"{o}}ink {et~al.}(2013)H{\"{o}}ink, Lenardic, \&
  Jellinek}]{Hoink2013}
H{\"{o}}ink, T., Lenardic, A., \& Jellinek, A.~M. 2013, {Earth's thermal
  evolution with multiple convection modes: A Monte-Carlo approach}, , ,
  doi:10.1016/j.pepi.2013.06.004

\bibitem[{Jackson \& Pollack(1984)}]{Jackson1984}
Jackson, M.~J., \& Pollack, H.~N. 1984, Journal of Geophysical Research, 89,
  10103

\bibitem[{Jannasch \& Mottl(1985)}]{Jannasch1985}
Jannasch, H.~W., \& Mottl, M.~J. 1985, Science, 229, 717

\bibitem[{Jaupart {et~al.}(2007)Jaupart, Labrosse, \& Mareschal}]{Jaupart2007}
Jaupart, C., Labrosse, S., \& Mareschal, J.~C. 2007 (Elsevier B.V.), 253--303,
  doi:10.1016/B978-044452748-6.00114-0

\bibitem[{Karato \& Wu(1993)}]{Karato771}
Karato, S.~I., \& Wu, P. 1993, Science, 260, 771

\bibitem[{Kennedy \& O'Hagan(2001)}]{Kennedy2001}
Kennedy, M.~C., \& O'Hagan, A. 2001, Journal of the Royal Statistical Society:
  Series B (Statistical Methodology), 63, 425

\bibitem[{Kite {et~al.}(2009)Kite, Manga, \& Gaidos}]{Kite2009}
Kite, E.~S., Manga, M., \& Gaidos, E. 2009, Astrophysical Journal, 700, 1732

\bibitem[{Komacek \& Abbot(2016)}]{Komacek2016}
Komacek, T.~D., \& Abbot, D.~S. 2016, The Astrophysical Journal, 832, 54

\bibitem[{Korenaga(2003)}]{Korenaga2003}
Korenaga, J. 2003, Geophysical Research Letters, 30, 47

\bibitem[{Korenaga(2006)}]{Korenaga2006}
---. 2006, in Geophysical Monograph Series, Vol. 164 (American Geophysical
  Union (AGU)), 7--32

\bibitem[{Korenaga(2008)}]{Korenaga2008}
---. 2008, {Urey ratio and the structure and evolution of Earth's mantle},
  John Wiley {\&} Sons, Ltd, doi:10.1029/2007RG000241

\bibitem[{Korenaga(2011)}]{Korenaga2011}
---. 2011, Journal of Geophysical Research: Solid Earth, 116, 20

\bibitem[{Korenaga(2016)}]{Korenaga2016}
---. 2016, Science Advances, 2, e1601168

\bibitem[{Lenardic(2018)}]{Lenardic}
Lenardic, A. 2018, Philosophical Transactions of the Royal Society A:
  Mathematical, Physical and Engineering Sciences, 376,
  doi:10.1098/rsta.2017.0416

\bibitem[{Lenardic {et~al.}(2011)Lenardic, Cooper, \& Moresi}]{Lenardic2011}
Lenardic, A., Cooper, C.~M., \& Moresi, L. 2011, Physics of the Earth and
  Planetary Interiors, 188, 127

\bibitem[{Lenardic \& Moresi(2003)}]{Lenardic2003}
Lenardic, A., \& Moresi, L. 2003, Physics of Fluids, 15, 455

\bibitem[{Loucks {et~al.}(2005)Loucks, van Beek, Stedinger, Dijkman, \&
  Villars}]{Loucks}
Loucks, D.~P., van Beek, E., Stedinger, J.~R., Dijkman, J.~P., \& Villars,
  M.~T. 2005, {Water Resources Systems Planning and Management and
  Applications: An Introduction to Methods, Models and Applications}, Vol.~51,
  57, arXiv:arXiv:1011.1669v3

\bibitem[{McKenzie \& Parker(1967)}]{McKenzie1967}
McKenzie, D.~P., \& Parker, R.~L. 1967, Nature, 216, 1276

\bibitem[{McNamara \& {Van Keken}(2000)}]{McNamara2000}
McNamara, A.~K., \& {Van Keken}, P.~E. 2000, Geochemistry, Geophysics,
  Geosystems, 1, n/a

\bibitem[{Moore \& Lenardic(2015)}]{Moore2015}
Moore, W.~B., \& Lenardic, A. 2015, Geophysical Research Letters, 42, 9255

\bibitem[{Moresi \& Solomatov(1998)}]{Moresi1998}
Moresi, L., \& Solomatov, V. 1998, Geophysical Journal International, 133, 669

\bibitem[{Morgan(1968)}]{Morgan1968}
Morgan, W.~J. 1968, Journal of Geophysical Research, 73, 1959

\bibitem[{{P. Fleming} \& VanderPlas(2018)}]{Fleming}
{P. Fleming}, D., \& VanderPlas, J. 2018, Journal of Open Source Software, 3,
  781

\bibitem[{Richards \& Lenardic(2018)}]{Richards2018}
Richards, M.~A., \& Lenardic, A. 2018, Geochemistry, Geophysics, Geosystems,
  19, 4858

\bibitem[{Rushby {et~al.}(2018)Rushby, Johnson, Mills, Watson, \&
  Claire}]{Rushby2018}
Rushby, A.~J., Johnson, M., Mills, B.~J., Watson, A.~J., \& Claire, M.~W. 2018,
  Astrobiology, 18, 469

\bibitem[{Saltelli(2019)}]{Saltelli2019}
Saltelli, A. 2019, Nature Communications, 10, 3870

\bibitem[{Schaefer \& Sasselov(2015)}]{Schaefer2015}
Schaefer, L., \& Sasselov, D. 2015, Astrophysical Journal, 801, 40

\bibitem[{Schubert \& Anderson(1985)}]{Schubert1985}
Schubert, G., \& Anderson, C.~A. 1985, Geophysical Journal of the Royal
  Astronomical Society, 80, 575

\bibitem[{Schubert {et~al.}(1979)Schubert, Cassen, \& Young}]{Schubert1979}
Schubert, G., Cassen, P., \& Young, R.~E. 1979, Icarus, 38, 192

\bibitem[{Schubert {et~al.}(1980)Schubert, Stevenson, \& Cassen}]{Schubert1980}
Schubert, G., Stevenson, D., \& Cassen, P. 1980, Journal of Geophysical
  Research, 85, 2531

\bibitem[{Seales {et~al.}(2019)Seales, Lenardic, \& Moore}]{Seales2019}
Seales, J., Lenardic, A., \& Moore, W.~B. 2019, Journal of Geophysical
  Research: Planets, 124, 2213

\bibitem[{Sleep(2000)}]{Sleep2000}
Sleep, N.~H. 2000, Journal of Geophysical Research E: Planets, 105, 17563

\bibitem[{Spohn \& Schubert(1982)}]{Spohn1982}
Spohn, T., \& Schubert, G. 1982, Journal of Geophysical Research, 87, 4682

\bibitem[{Taleb(2012)}]{Taleb2012}
Taleb, N.~N. 2012, SSRN Electronic Journal, arXiv:1209.2298

\bibitem[{Tosi {et~al.}(2017)Tosi, Godolt, Stracke, Ruedas, Grenfell,
  H{\"{o}}ning, Nikolaou, Plesa, Breuer, \& Spohn}]{Tosi2017}
Tosi, N., Godolt, M., Stracke, B., {et~al.} 2017, Astronomy and Astrophysics,
  605, arXiv:1707.06051

\bibitem[{Tozer(1972)}]{Tozer1972}
Tozer, D.~C. 1972, Physics of the Earth and Planetary Interiors, 6, 182

\bibitem[{Turcotte {et~al.}(2002)Turcotte, Schubert, \&
  Turcotte}]{Turcotte2002}
Turcotte, D.~L., Schubert, G., \& Turcotte, D.~L. 2002, {Geodynamics}, 2nd edn.
  (Cambridge: Cambridge University Press), 456

\bibitem[{Zhong {et~al.}(2000)Zhong, Zuber, Moresi, \& Gurnis}]{Zhong2000}
Zhong, S., Zuber, M.~T., Moresi, L., \& Gurnis, M. 2000, Journal of Geophysical
  Research: Solid Earth, 105, 11063

\end{thebibliography}
\bibliographystyle{aasjournal}



\end{document}